%% file: LO_Arch_Opt.tex
\documentclass[journal]{IEEEtran}
\ifCLASSINFOpdf
  \usepackage[pdftex]{graphicx}
  \graphicspath{{./images/}}
\else
\fi
\usepackage{amsmath}
\usepackage{array}
\usepackage[caption=false,font=footnotesize]{subfig}
\bstctlcite{ctrl_params}
\usepackage{cite}

\begin{document}

%
% paper title
% Titles are generally capitalized except for words such as a, an, and, as,
% at, but, by, for, in, nor, of, on, or, the, to and up, which are usually
% not capitalized unless they are the first or last word of the title.
% Linebreaks \\ can be used within to get better formatting as desired.
% Do not put math or special symbols in the title.
\title{Optimizing the LO Distribution Architecture of mm-Wave Massive MIMO Receivers}
%
%
% author names and IEEE memberships
% note positions of commas and nonbreaking spaces ( ~ ) LaTeX will not break
% a structure at a ~ so this keeps an author's name from being broken across
% two lines.
% use \thanks{} to gain access to the first footnote area
% a separate \thanks must be used for each paragraph as LaTeX2e's \thanks
% was not built to handle multiple paragraphs
%

\author{Greg~LaCaille,
		Antonio~Puglielli,
        Elad~Alon,
        Borivoje~Nikolic,
        Ali~Niknejad,% <-this % stops a space
\thanks{ G. LaCaille, A. Puglielli, E. Alon, B. Nikolic, and A. Niknejad are with the Department of Electrical Engineering and Computer Sciences, University of California at Berkeley, Berkeley, CA 94720 USA}

\thanks{ This work was supported by NSF EARS ECCS-1642920, NSF ECCS-1232318, and Intel Corporation.}}

\maketitle

% As a general rule, do not put math, special symbols or citations
% in the abstract or keywords.
\begin{abstract}
Wireless networks at millimeter wavelengths have significant implementation difficulties. The path loss at these frequencies naturally leads us to consider antenna arrays with many elements. In these arrays, local oscillator (LO) generation is particularly challenging since the LO specifications affect the system architecture, signal processing design, and circuit implementation. We thoroughly analyze the effect of LO architecture design choices on the performance of a mm-wave massive MIMO uplink. This investigation focuses on the tradeoffs involved in centralized and distributed LO generation, correlated and uncorrelated phase noise sources, and the bandwidths of PLLs and carrier recovery loops. We show that, from both a performance and implementation complexity standpoint, the optimal LO architecture uses several distributed subarrays locked to a single intermediate-frequency reference in the low GHz range. Additionally, we show that the choice of PLL and carrier recovery loop bandwidths strongly affects the performance; for typical system parameters, loop bandwidths on the order of tens of MHz achieve SINRs suitable for high-order constellations. Finally, we present system simulations incorporating a complete model of the LO generation system and consider the case of a 128-element array with 16x-spatial multiplexing and a 2 GHz channel bandwidth at 75 GHz carrier. Using our optimization procedure we show that the system can support 16-way spatial multiplexing with 64-QAM modulation.    
\end{abstract}

% Note that keywords are not normally used for peerreview papers.

% For peer review papers, you can put extra information on the cover
% page as needed:
% \ifCLASSOPTIONpeerreview
% \begin{center} \bfseries EDICS Category: 3-BBND \end{center}
% \fi
%
% For peerreview papers, this IEEEtran command inserts a page break and
% creates the second title. It will be ignored for other modes.
\IEEEpeerreviewmaketitle

\input{sec-Intro}
\input{sec-LO_dist_choices}
\input{sec-CR_PLL_single_element}
\input{sec-single_element}

\input{sec-single_user_beam_scramble}
\input{sec-multi_user}

\input{sec-conclusion}
\input{sec-appendix}
\input{sec-Acknow}

\ifCLASSOPTIONcaptionsoff
  \newpage
\fi

% trigger a \newpage just before the given reference
% number - used to balance the columns on the last page
% adjust value as needed - may need to be readjusted if
% the document is modified later
%\IEEEtriggeratref{8}
% The "triggered" command can be changed if desired:
%\IEEEtriggercmd{\enlargethispage{-5in}}

% references section

% can use a bibliography generated by BibTeX as a .bbl file
% BibTeX documentation can be easily obtained at:
% http://mirror.ctan.org/biblio/bibtex/contrib/doc/
% The IEEEtran BibTeX style support page is at:
% http://www.michaelshell.org/tex/ieeetran/bibtex/
%\bibliographystyle{IEEEtran}
% argument is your BibTeX string definitions and bibliography database(s)
%\bibliography{IEEEabrv,../bib/paper}
%
% <OR> manually copy in the resultant .bbl file
% set second argument of \begin to the number of references
% (used to reserve space for the reference number labels box)

\input{sec-bib}

% biography section
% 
% If you have an EPS/PDF photo (graphicx package needed) extra braces are
% needed around the contents of the optional argument to biography to prevent
% the LaTeX parser from getting confused when it sees the complicated
% \includegraphics command within an optional argument. (You could create
% your own custom macro containing the \includegraphics command to make things
% simpler here.)
%\begin{IEEEbiography}[{\includegraphics[width=1in,height=1.25in,clip,keepaspectratio]{mshell}}]{Michael Shell}
% or if you just want to reserve a space for a photo:

%\begin{IEEEbiography}{Greg LaCaille}
%Biography text here.
%\end{IEEEbiography}

%\begin{IEEEbiography}{Antonio Puglielli}
%Biography text here.
%\end{IEEEbiography}

%\begin{IEEEbiography}{Gregory Wright}
%Biography text here.
%\end{IEEEbiography}

%\begin{IEEEbiography}{Elad Alon}
%Biography text here.
%\end{IEEEbiography}

%\begin{IEEEbiography}{Borivoje Nikolic}
%Biography text here.
%\end{IEEEbiography}

%\begin{IEEEbiography}{Ali Niknejad}
%Biography text here.
%\end{IEEEbiography}

% insert where needed to balance the two columns on the last page with
% biographies
%\newpage

% You can push biographies down or up by placing
% a \vfill before or after them. The appropriate
% use of \vfill depends on what kind of text is
% on the last page and whether or not the columns
% are being equalized.

%\vfill

% Can be used to pull up biographies so that the bottom of the last one
% is flush with the other column.
%\enlargethispage{-5in}

% that's all folks
\end{document}

%% file: sec-Intro.tex
\section{Introduction}
% The very first letter is a 2 line initial drop letter followed
% by the rest of the first word in caps.
% 
% form to use if the first word consists of a single letter:
% \IEEEPARstart{A}{demo} file is ....
% 
% form to use if you need the single drop letter followed by
% normal text (unknown if ever used by the IEEE):
% \IEEEPARstart{A}{}demo file is ....
% 
% Some journals put the first two words in caps:
% \IEEEPARstart{T}{his demo} file is ....
% 
% Here we have the typical use of a "T" for an initial drop letter
% and "HIS" in caps to complete the first word.
\IEEEPARstart{C}{ellular} wireless networks are expected to enter their fifth generation (5G) around 2020.  Massive MIMO and use of mm-wave bands are both expected to be cornerstone technologies for 5G wireless networks \cite{Andrews:2014, Rappaport:2013}. These next generation networks seek to offer revolutionary performance, diminishing the need for wired connectivity to the billions of devices envisioned in the Internet of Everything (IoE). It is expected that 5G will offer a 1000x increase in cell capacity, with sub-10ms latencies and increased reliability \cite{Andrews:2014}. Massive MIMO promises to address many of these requirements by enabling a huge increase in spectral efficiencies \cite{Marzetta:2010}, while mm-wave frequencies unlock much larger channel bandwidths than are available in current RF systems \cite{Rappaport:2013}.

5G wireless communication systems at mm-wave present a significant design challenge. Advances in semiconductor technology (i.e. Moore's law) are usually led by improvements in digital CMOS, where the focus is on the density and speed of logic/memory devices. While these technology advancements help some aspects of wireless transceiver design, for example, through improvement in $f_t$, other parameters that affect analog performance, such as $f_{max}$ and $g_mr_o$, improve at a much slower rate and may even degrade in deeply scaled technologies \cite{Niknejad:2015}. This means that while implementing high-datarate transceivers at mm-wave is possible, it requires very careful system architecture and system-circuit co-design in order to achieve satisfactory system performance. 

Recently described mm-wave chipsets are mainly targeted at the 802.11ad WLAN standard in the unlicensed 60 GHz band \cite{IEEE802:ad}. These implementations generally consist of 16- or 32-element arrays, with a range of around one to twenty meters for indoor, low-mobility scenarios. These systems provide single-user communication over a bandwidth of 2 GHz with single-carrier QPSK or 16-QAM constellations \cite{Boers:2014,Saito:2013,Okada:2013}. 
In contrast, 5G mm-wave communication intends to support ranges beyond 100 meters, aggressive multi-user spatial multiplexing, and higher user mobility \cite{Andrews:2014, Alkhateeb:2014, Puglielli_proc:2016}. This will require the use of significantly larger arrays (in excess of 100 elements) while still meeting challenging power and cost budgets. 

Of all the components of a base-station, the LO generation system is most strongly coupled to system architecture, signal processing, and circuit implementation. First, because the LO determines the phase and frequency synchronization of the array its quality has a significant bearing on the design of beamforming and carrier recovery algorithms. Second, because the LO must be routed to every single element, the cost of LO distribution, measured in power and complexity, affects the overall system architecture. Finally, the overall system performance will depend on the circuit specifications and implementation. As a result, array architecture, signal processing algorithms, and circuits must be carefully co-designed in order to manage complexity and ensure performance throughout the LO generation system.  

A number of previous works have analyzed the effect of phase noise in massive MIMO systems \cite{Pitarokoilis2015-wp,Pitarokoilis2016-rd,Pitarokoilis2013-hp,Pitarokoilis2016-dw,Mehrpouyan2012-zj,Krishnan2016-xc,Krishnan2015-yk,Khanzadi2015-nv,Hohne2010-yg,Gustavsson2014-jq,Durisi2014-ad,Bjornson2015-eb,Bjornson2015-xl,Puglielli_icc:2016}. In particular, Pitarokoilis conducted a thorough analysis of a single-carrier massive MIMO system with either fully synchronous or fully asynchronous oscillators \cite{Pitarokoilis2015-wp}. Krishnan has proposed techniques for tracking and reducing the effect of phase noise \cite{Krishnan2015-yk}. Finally, Bj{\"o}rnson has considered how hardware and communication link design inform each other \cite{Bjornson2015-xl}. 

To the best of our knowledge, few works consider how phase-locked loops (PLLs) are used in mm-wave arrays. Thomas analyzed the performance of a SISO 72 GHz mm-wave system with a realizable phase noise specification and considered the performance of a squaring-loop carrier recovery scheme \cite{Thomas:2015}. Puglielli analyzed the effect of PLLs in massive MIMO systems using OFDM \cite{Puglielli_icc:2016}. In this work we consider the performance of mm-wave massive MIMO arrays using practical LO generation schemes. The local oscillator PLLs introduce phase noise correlations among elements and both the beamforming and carrier recovery loops are influenced by the spectral distribution of the phase noise. This gives us the opportunity to make important design trade-offs that are often not accounted for. We compare several different system architectures for mm-wave arrays, identify the key parameters and trade-offs, and derive specific guidelines for the design of the PLLs and carrier recovery. Our system simulations show that with the proposed improvements, mm-wave systems can reach the level of performance to support multi-user spatial multiplexing with high-order constellations in a energy efficient manner. 

\section{mm-Wave Channel Propagation}
	Propagation at mm-wave is notably different than at lower frequencies. The most significant effect is that a mm-wave antenna with the same pattern as a lower frequency antenna captures much less of the incident energy flux. If the physical dimensions of the antenna are scaled to give a constant pattern, the energy captured drops like $\lambda^{-2}$. The decrease of energy captured with increasing frequency between antennas of identical patterns is described by the Friis equation and is usually called ``path loss'', although it is not a property of the path \cite{Friis:1946}. In addition to this, many surfaces exhibit roughness on length scales comparable to mm wavelength. This causes scattering to be diffusive rather than specular, again increasing the path loss \cite{Kyro:2013}.
    
    The larger channel bandwidths targeted for mm-wave also influence the link budget. As channel bandwidth increases, the power spectral density of the noise floor remains constant, while the power spectral density of the transmitter reduces for a fixed total output power, leading to a reduction in SNR. If the range of a cell at mm-wave is to be comparable to an RF cell, the increase in path loss and decrease in transmit power spectral density must be compensated to achieve the same SNR. One possibility to overcome the potential reduction in SNR is to increase the transmit power; however, this is undesirable as RF power amplifiers already operate at an output power optimal for typical supply voltages and antenna impedance\cite{Niknejad_pa:2012}. Increasing the transmit power will also lead to a greater drain on the battery of mobile devices. A more attractive choice is to make up the link budget gap by increasing the antenna gain, exploiting directivity to increase the power in the intended direction without an additional cost in total radiated power. Table \ref{table:link_budget} shows typical RF and mm-wave link budgets using both transmit power and antenna gain meet the increased requirements. The loss exponents are from empirical measurements of small cell scale propagation environments \cite{Durgin:1998,Rappaport:2013_channel}. To meet the mm-wave link budget without any additional antenna gain, the PA would need to radiate over a kilowatt.  However, using fixed directional antennas (e.g., horns) or arrays of 128 and 16 elements at the base station and handset respectively, the link budget could be satisfied with a far more reasonable transmit power.

	Because mm-wave system typically use directional antennas and also because of the higher loss of scattered multipath rays, mm-wave channels are often modeled as Rician with strong line of sight components. In this situation, single carrier modulation is often preferred to OFDM as it does not require transmitting signals with high peak-to-average power ratio \cite{Kato:2009}. Recent testbeds operating at E-band have used single-carrier modulation for this reason \cite{Cudak:2014}. In what follows, we have modeled our system using a single carrier modulation scheme over line of sight (LOS) channels.

\newcolumntype{L}[1]{>{\raggedright\let\newline\\\arraybackslash\hspace{0pt}}m{#1}}
\newcolumntype{C}[1]{>{\centering\let\newline\\\arraybackslash\hspace{0pt}}m{#1}}
\newcolumntype{R}[1]{>{\raggedleft\let\newline\\\arraybackslash\hspace{0pt}}m{#1}}

\begin{table}[!t]
%% increase table row spacing, adjust to taste
\renewcommand{\arraystretch}{1.3}
% if using array.sty, it might be a good idea to tweak the value of
%\extrarowheight as needed to properly center the text within the cells
\caption{RF and mm-wave link budgets}
\label{table:link_budget}
\centering
%% Some packages, such as MDW tools, offer better commands for making tables
%% than the plain LaTeX2e tabular which is used here.
\begin{tabular}{|C{1.2in}|C{0.4in}|C{0.4in}|C{0.4in}|}
\hline
	 								& RF 	& mmw \#1 	& mmw \#2 	\\
\hline
Bandwidth (MHz) 					& 20 	& 2000			& 2000			\\
\hline
Rx NF (dB) 							& 5.0	& 5.0			& 5.0			\\
\hline
Input Referred Noise Power (dBm) 	& -96.0 & -76.0			& -76.0 		\\
\hline
Carrier Freq (GHz) 					& 2.5	& 60			& 60			\\
\hline
Loss Exponent 						& 2.9	& 2.2			& 2.2			\\
\hline
Loss @ 100m LOS (dB) 				& 98.4	& 112.0			& 112.0			\\
\hline
Rx SNR (dB) 						& 26.0	& 26.0			& 26.0			\\
\hline
BS Antenna Gain (dB) 				& 0.0	& 0.0 			& 21.0			\\
\hline
UE Antenna Gain (dB) 				& 0.0	& 0.0 			& 12.0			\\
\hline
UE Radiated \newline Tx Power (dBm) & 28.4 	& 62.0			& 28.9			\\
\hline
\end{tabular}
\end{table}

%% file: sec-LO_dist_choices.tex
\begin{figure*}[!t]
\centering
\subfloat[Central carrier generation (CCG)]{\includegraphics[width=2in]{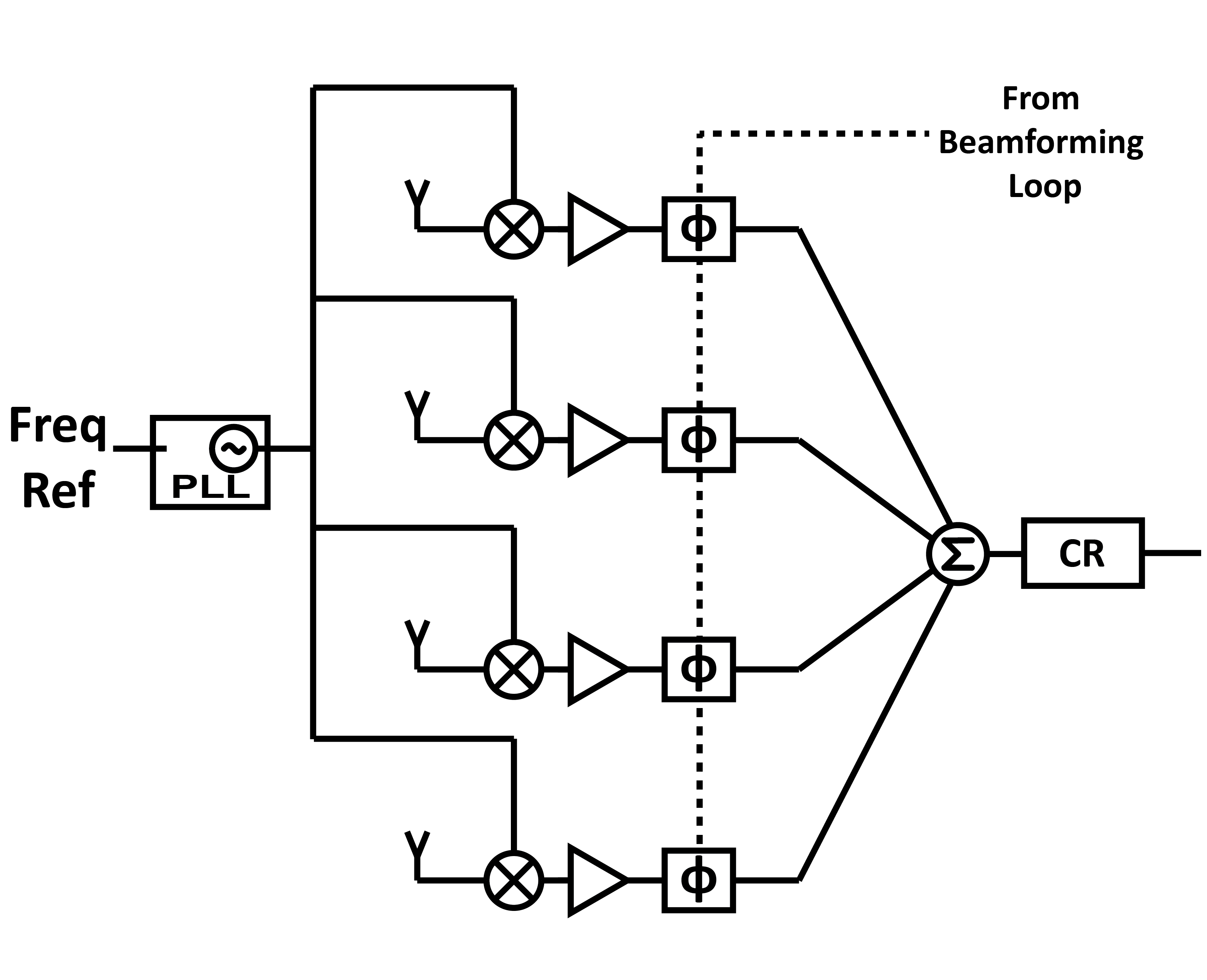}
\label{fig:central_block}}
\hfil
\subfloat[Local carrier generation (LCG)]{\includegraphics[width=2in]{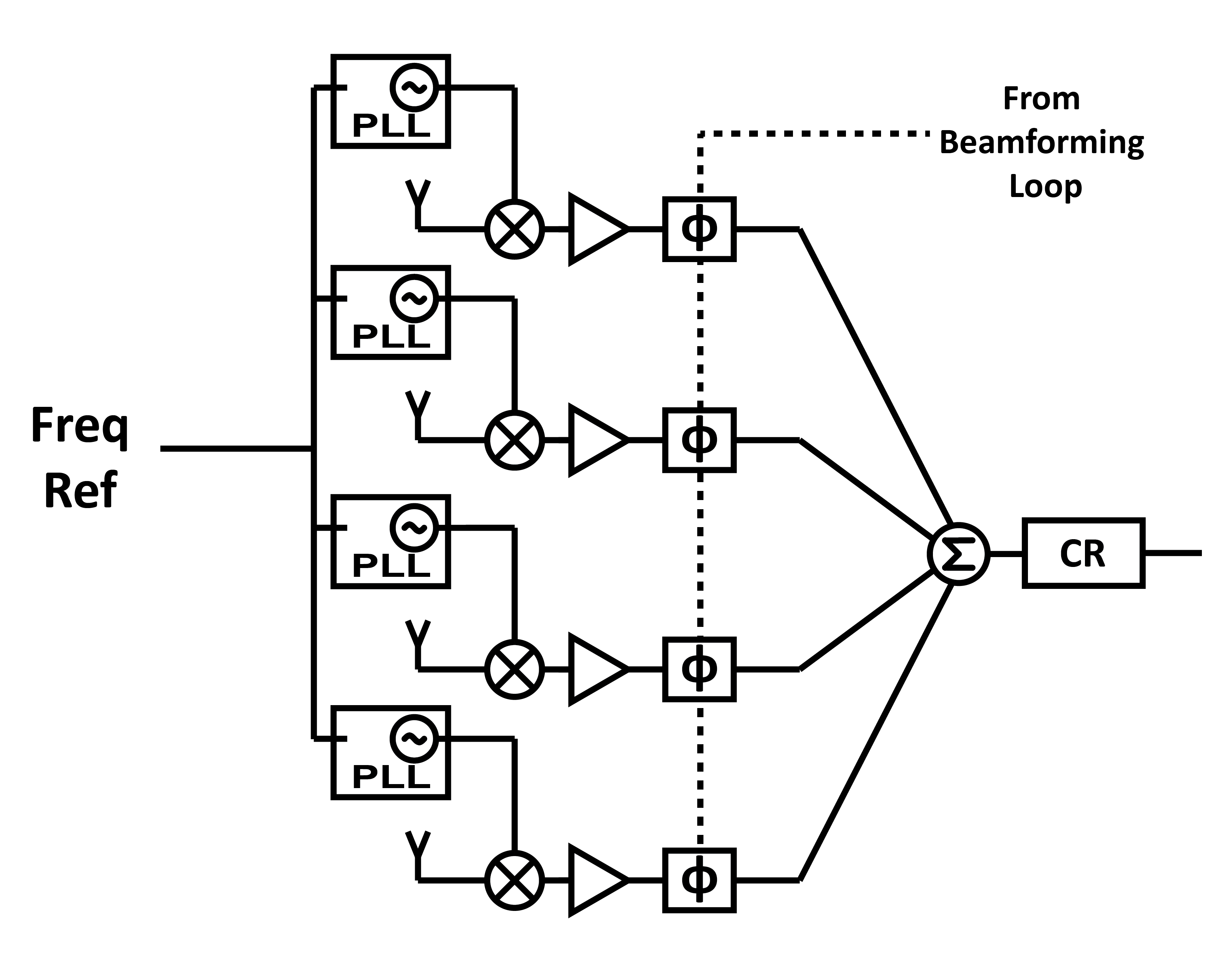}
\label{fig:local_block}}
\hfil
\subfloat[Generalized carrier generation (GCG)]{\includegraphics[width=2in]{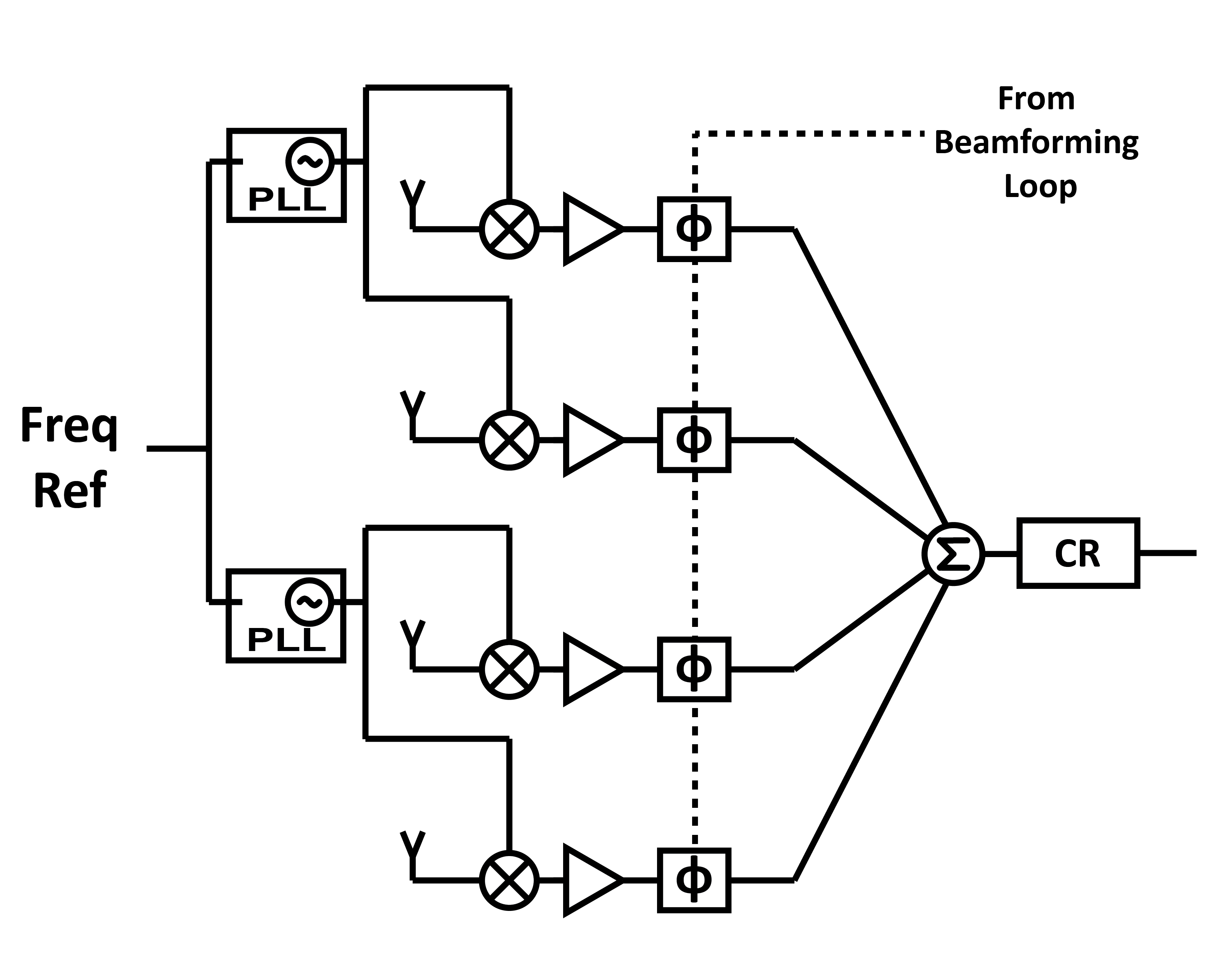}
\label{fig:segmented_block}}

\caption{Example block diagrams for different LO distribution schemes.}
\label{fig:central_v_local_block}
\end{figure*}

\section{LO Distribution for Large Scale Arrays}
	Distributing a mm-wave LO to an array consisting of hundreds of elements can potentially consume a large fraction of the total power in the array. The power consumption will be affected by both circuit design and system-level architecture choices and both must be considered together to achieve optimal performance.
    
	The power consumption of the entire LO chain can be expressed as:
\begin{equation}
P_{LO} = P_{load} + P_{distr} + P_{VCO} + P_{PLL} + P_{ref}.
\end{equation}
Here $P_{load}$ represents the power that must be delivered to the load (e.g. the mixers), $P_{distr}$ is the power required to route the high-frequency LO, $P_{VCO}$ is the power burned in the VCO, $P_{PLL}$ is the power burned in the rest of the PLL, and $P_{ref}$ is the power for reference distribution. 

	Figure \ref{fig:central_v_local_block} presents several different LO distribution architectures. A conceptually simple option is a central carrier generation (CCG) scheme, where a single central PLL generates the LO with the desired phase noise profile and that LO is distributed at the carrier frequency to each element (Figure \ref{fig:central_block}). In this scenario, the LO distribution network needs enough gain to overcome the large loss associated with routing over a distance of many wavelengths and splitting to each element; the distribution buffers would therefore burn a large amount of power.

The opposite option is a local carrier generation (LCG) scheme, where a local PLL is used at each element and all the PLLs are frequency locked to a common low-frequency reference (Figure \ref{fig:local_block}). In this scenario the power consumption is dominated by the hundreds of PLLs since the mm-wave distribution is very short and the reference distribution consumes a trivial amount of power. At first it may seem that building a phased array with hundreds of PLLs would inherently be inefficient; however, by exploiting the averaging of the uncorrelated noise of individual oscillators, to first order the combined power of the VCOs in LCG should equal the power of the VCO in the CCG scheme \cite{Puglielli_icc:2016}.

The trade-off between the CCG and LCG schemes comes from balancing distribution power against PLL overhead power. In practice we can consider a generalized carrier generation (GCG) scheme, consisting of several PLLs that each serve multiple elements (Figure \ref{fig:segmented_block}). In the following section we present a detailed analysis of how the total power dissipation in GCG scales as a function of the number of elements per PLL.

\subsection{VCO Power Consumption and Array Gain}

In general, $P_{VCO} > P_{PLL}$, and for a given VCO FoM, which is set by the quality factor and circuit architecture, the amount of VCO phase noise is inversely proportional to the VCO's power consumption \cite{Murphy:2010, Hajimiri:1999, Bank:2006}.
\begin{equation}
\mathcal{L}(f_{\Delta}) = \frac{(\frac{f_{LO}}{f_{\Delta}})^2 FoM_{VCO}}{P_{VCO}}
\label{eq:vco-fom}
\end{equation}
Therefore the VCO power is largely determined by the allowable phase noise budget. 

It has been shown that in arrays, the uncorrelated phase noise of individual VCOs is averaged to yield an improvement proportional to the number of oscillators \cite{Hohne2010-yg, Puglielli_icc:2016, Pitarokoilis2015-wp}. This means that, compared with a single VCO, an array of $N$ VCOs can relax the performance of each individual VCO by $N$; consequently, each VCO's power can be relaxed by the same factor. As a result, to first order the power-performance point of an array of VCOs is independent of the number of VCOs. 

\subsection{Comparison of Distribution Architectures}

\begin{figure}[!t]
\centering
\includegraphics[width=3.5in]{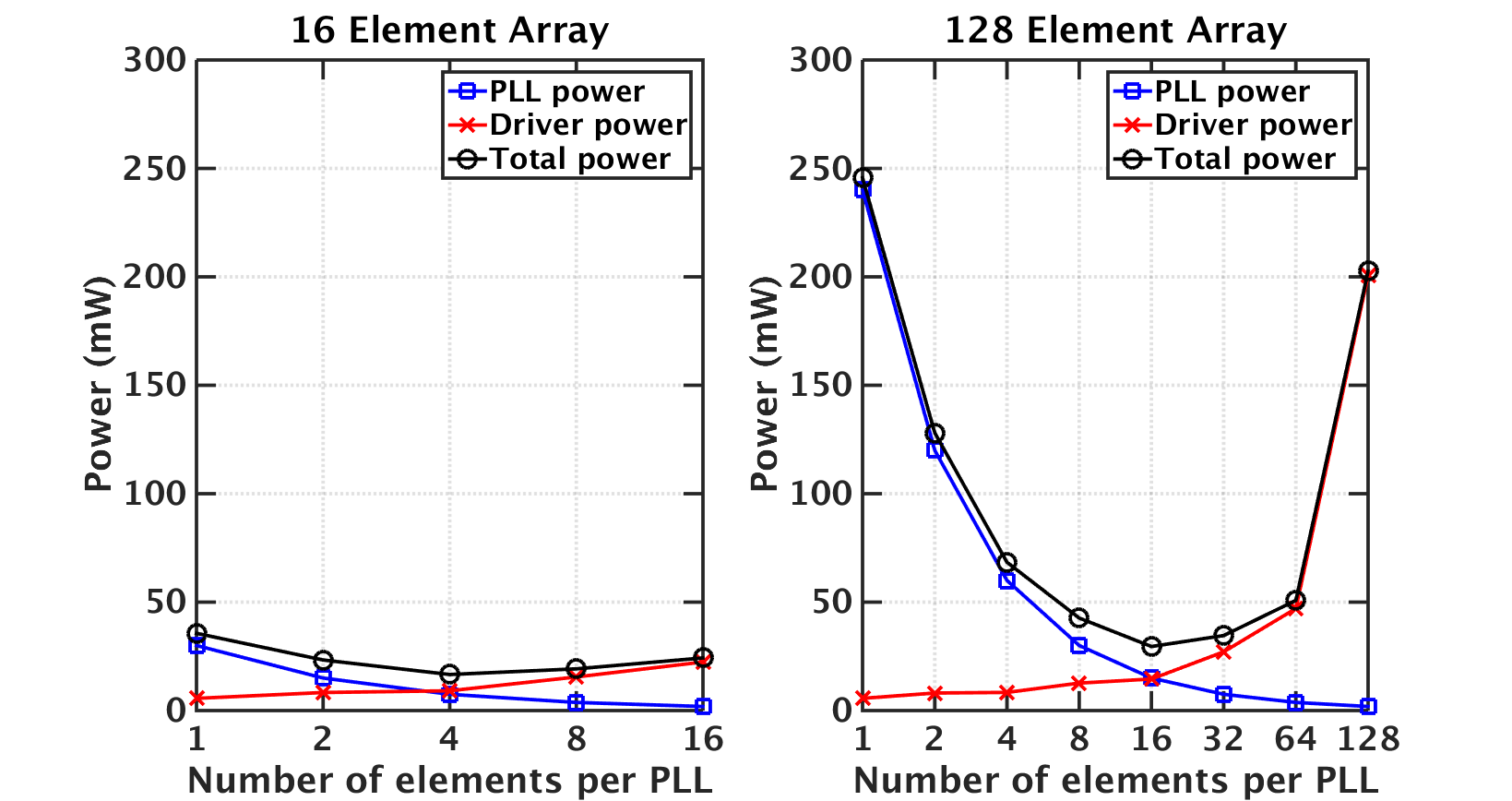}
\caption{LO distribution power consumption as a function of number of elements per PLL, for 16- and 128-element arrays. Model parameters are taken from recent literature or products, as described in the Appendix.}
\label{fig:routing-loss}
\end{figure}  
    
    The power of different LO distribution architectures can be compared quantitatively as a function of $N$, the number of elements per PLL. $N=1$ corresponds to the LCG scheme and $N=M$ to the CCG scheme, where $M$ is the number of array elements. To perform this analysis it is necessary to study how each power contribution depends on $N$. It is clear that the load power is independent of $N$. The VCO power can be made independent of $N$ by exploiting array averaging. Finally, the reference distribution power can be neglected since it is very small\footnote{The power is given by $\frac{1}{2}CV^2f$ where $C$ is the capacitance, $V$ the voltage, and $f$ the frequency. Since the array is physically very small, the capacitance is on the order of picofarads and the reference frequency is only 100 MHz, giving negligible power consumption.}. Therefore, $P_{distr}$ and $P_{PLL}$ are most closely tied to the overall system architecture. $P_{distr}$ accounts for the power needed to overcome loss in the distribution network, which primarily consists of wire loss and the loss in power splitters\footnote{This accounts for \textit{excess loss} above the desired power splitting.}. It is important to note that these losses increase significantly at high frequencies. $P_{PLL}$ depends on both the design of the PLL as well as the architectural choice of how many PLLs to use.

	The Appendix presents a detailed model of the LO power consumption accounting for routing loss, splitter loss, and number of PLLs. Using this model, Figure \ref{fig:routing-loss} shows the power consumption as a function of $N$ (see Appendix for discussion of the assumptions) for both a 16- and 128-element array. With only 16 elements the difference between architectures is minimal. In contrast, for a 128-element array, the choice of architecture is very significant to the overall power and complexity of the LO chain. By choosing the optimum architecture, the power can be reduced by 5-10x compared to a CCG or LCG implementation. 
    
    This analysis illustrates that for the large arrays required in 5G mm-wave systems, the choice of LO architecture is very impactful. In particular, the optimum architecture involves distributed generation of the LO which potentially runs the risk of de-synchronizing the elements. In light of these architectural optimizations, it is important to analyze how the choice of LO architecture influences the performance of the overall system. How does the high carrier frequency at mm-wave influence the phase noise budget? How does distributed frequency generation influence inter-element and inter-user synchronization? The remainder of the paper will study how the LO generation architecture affects the achievable performance of 5G mm-wave systems. We will use the CCG and LCG schemes to illustrate extremes of behavior and finally show that with multi-user operation, the performance of a GCG scheme strongly depends on the choice of $N$. 

%% file: sec-CR_PLL_single_element.tex
\section{Phase Noise Filtering Techniques}
Modern wireless systems rely on a wide variety of analog circuits, signal processing blocks, and control loops to achieve high fidelity data transmission. The overall performance degradation of the receiver from phase noise will be set by the interaction of several components in the system. 
    
\subsection{Phase locked loops (PLLs)}
    
\begin{figure}[!t]
\centering
\subfloat[]{\includegraphics[width=3.0in]{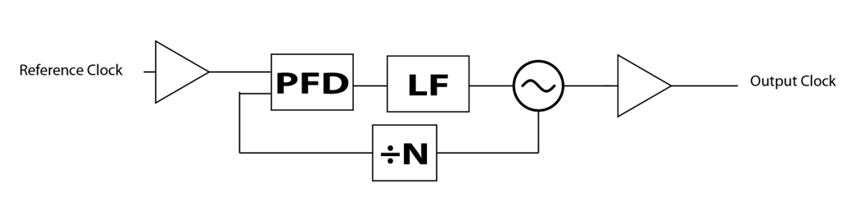}
\label{fig:pll_block}}

\subfloat[]{\includegraphics[width=3.2in]{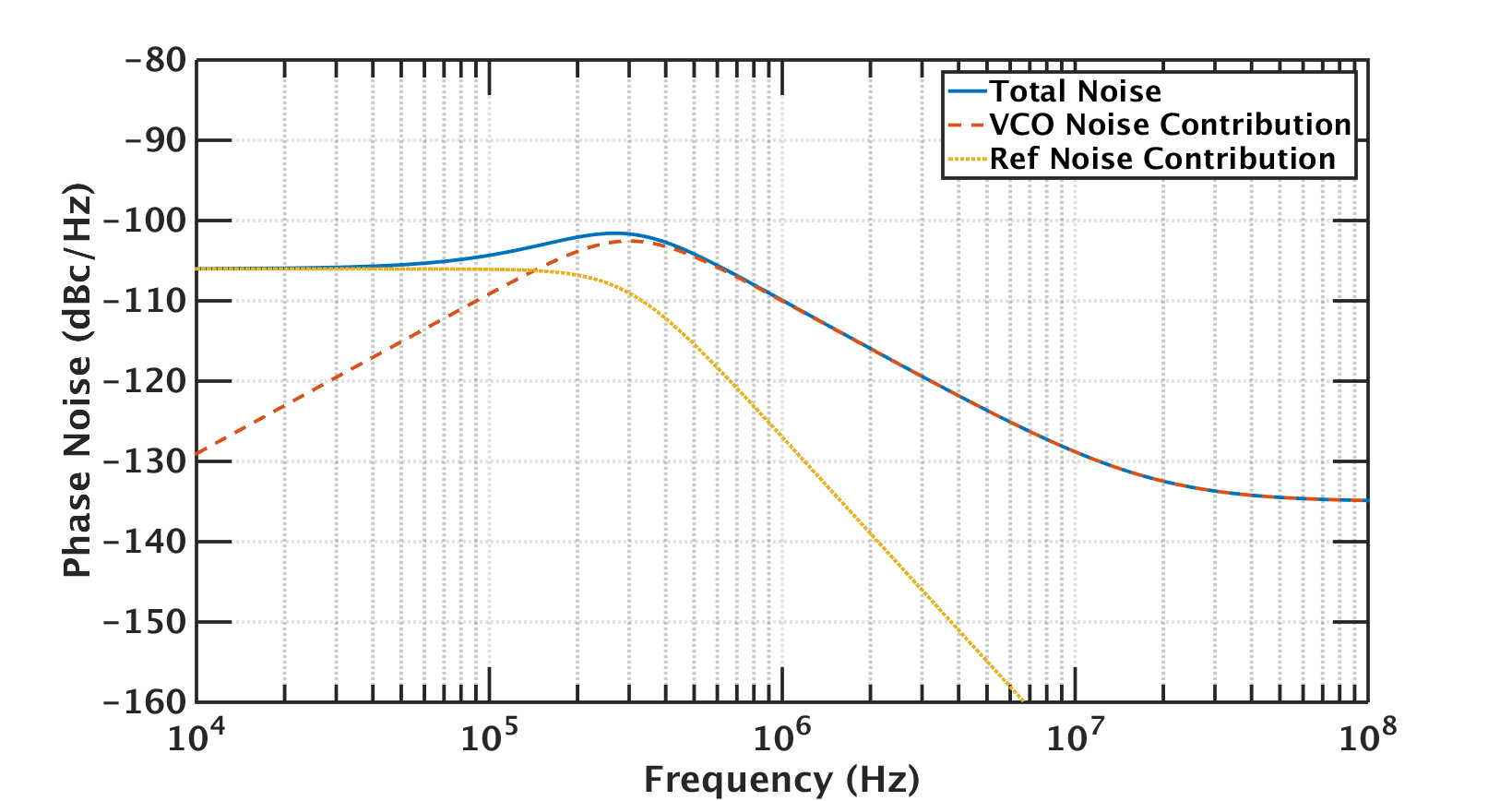}
\label{fig:pll_psd}}
\caption{a) Block diagram of standard PLL. b) Phase noise PSDs at the PLL output.}
\label{fig:pll_basics}
\end{figure}

	The LO is usually generated from a phase locked loop (PLL), which locks the phase of a voltage controlled oscillator (VCO) to a stable low-frequency reference. In this way, the PLL can achieve the frequency and phase accuracy of the reference while retaining the tunability of the VCO. A simplified block diagram of a PLL is shown in Figure \ref{fig:pll_block}. Typically the reference comes from a high quality crystal oscillator, which achieves excellent phase noise performance at a fixed frequency due to its very high Q. In fact, the phase noise on the reference is typically dominated by the thermal jitter introduced in its distribution buffers and the reference phase noise can be modeled simply with a white power spectral density (PSD). The VCO in the PLL will have its own phase noise, which can be modeled primarily as a $1/f^2$ PSD arising from a Wiener process \cite{Petrovic:2007}. Figure \ref{fig:pll_psd} shows the PSD of a PLL's noise and its different contributions.
    
%\begin{figure*}[!t]
%\centering
%\subfloat[Case I]{\includegraphics[width=2.5in]{box}%
%\label{fig_first_case}}
%\hfil
%\subfloat[Case II]{\includegraphics[width=2.5in]{box}%
%\label{fig_second_case}}
%\caption{Simulation results for the network.}
%\label{fig_sim}
%\end{figure*}
    
	The loop filter acts to low-pass filter the reference phase noise while high-pass filtering the VCO phase noise \cite{Hanumolu:2004}.    
\begin{equation}
\begin{split}
H_{ref}(s) = &\frac{f_{LO}}{f_{ref}}\frac{H_{LF}(s)}{1+H_{LF}(s)} \\
H_{vco}(s) &= \frac{1}{1+H_{LF}(s)}
\end{split}
\end{equation}
These transfer functions are typically second order, but significant design intuition can be obtained by assuming that all phase noise below the PLL bandwidth comes from the reference and all noise above the PLL bandwidth comes from the VCO. Since the reference and VCO noise have different PSDs, a trade-off in the integrated phase error as a function of PLL bandwidth exists. To first order, the optimum PLL bandwidth is the frequency at which the reference and VCO noise contribute equal amounts of noise to the output.

\begin{figure}[!t]
\centering
\includegraphics[width=3in]{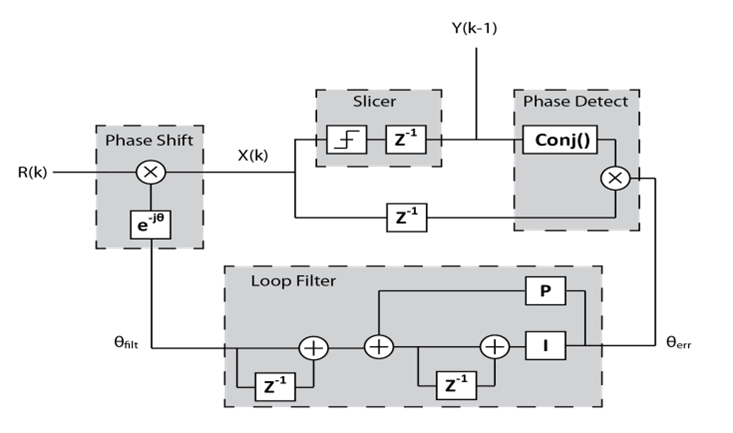}
\caption{Block diagram of decision-directed carrier recovery loop.}
\label{fig:cr_block}
\end{figure}

\subsection{Carrier recovery}
	
    The PLL is designed for optimal phase and frequency stability but is ultimately limited by the quality of the reference and circuits which are a function of technology limits. Therefore, carrier recovery (CR) loops are used to compensate for residual carrier phase and frequency offsets. Many implementations of CR with varying complexity and performance trade-offs exist \cite{Meyr:1998, Mengali:1997}. One example of a CR loop is the decision-directed PLL shown in Figure \ref{fig:cr_block}. Once the loop is locked, the phase of the measured symbol is compared to the true constellation point to compute the instantaneous phase error. This error signal is fed back through a loop filter to apply a correction to the signal path. The loop filter typically contains two poles at DC to eliminate static phase and static frequency errors. 
    
    This operation is essentially identical to the PLL of Figure \ref{fig:pll_basics} where the "reference" is the sequence of ideal constellation points. As a result, the CR loop applies a high-pass transfer function to the input phase noise. If the CR bandwidth is set very low, the loop will only remove static phase and frequency offsets. However, if the bandwidth of the CR is allowed to be large it can also track and filter out the instantaneous phase error arising from high-frequency components of the phase noise.

\subsection{Channel Estimation}
	In practical scenarios the wireless channel varies over time due to physical motion in the surrounding environment. To track these changes, a channel estimation procedure is run at a rate comparable to the channel coherence time. In array receivers, channel estimation is performed on a per-element basis to estimate the full channel matrix and apply beamforming. Since variations in LO phase are indistinguishable from variations in the phase of the wireless channel, per-element channel estimation is equivalent to a per-element CR operating at the channel estimation rate. As such, the channel estimation loop sets an absolute lower bound on the frequency at which phase noise will impact the receiver. In this work we focus on low mobility scenarios, where channel estimation is performed every 100 $\mu s$.

%% file: sec-single_element.tex
\begin{figure}[!t]
\centering
\includegraphics[width=3.5in]{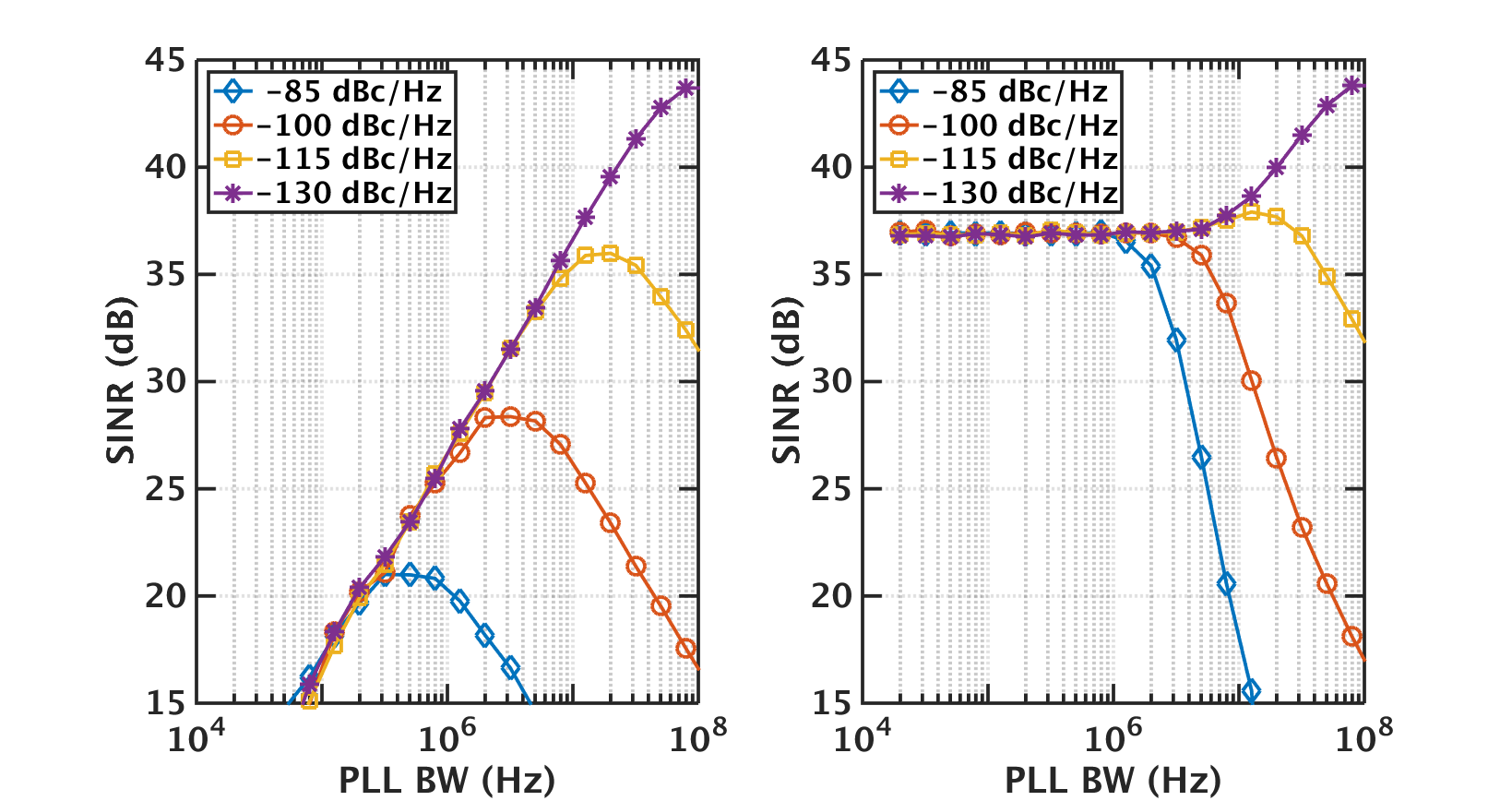}
\caption{Single-user SINR vs PLL bandwidth for various reference noise levels. Left panel has 10kHz carrier recovery bandwidth, right panel has 10 MHz carrier recovery bandwidth.}
\label{fig:single_user_pll_bw}
\end{figure}

\section{Reference Noise Elimination Using Carrier Recovery}
    
    Since the CR can filter phase noise, it can have a substantial impact on the achievable SINR of the system. Figure \ref{fig:single_user_pll_bw} shows the result from time-domain simulations of a receiver operating at 2 GSymbol/s for various levels of output-referred reference phase noise. The VCO noise is -90 dBc/Hz at 1 MHz offset in all cases. In Figure \ref{fig:single_user_pll_bw}a, 10 kHz CR bandwidth is used and consequently the SINR performance depends solely on the PLL bandwidth. As expected, there is a clear optimum PLL bandwidth for each level of reference noise, and as the reference quality is improved, the optimal PLL bandwidth increases. 
	
    In contrast, figure \ref{fig:single_user_pll_bw}b depicts the results when the CR bandwidth is increased to 10 MHz. This 10 MHz bandwidth is still 200x below the symbol rate and does not pose any stability problems for the second-order decision-directed loop. In this situation, the SINR is significantly improved for low PLL bandwidths. This is because at these low PLL bandwidths the phase error from the PLL is dominated by low frequency VCO noise. As long as this VCO noise is below the CR bandwidth it can be filtered by the CR loop. Once the PLL bandwidth approaches or exceeds the CR bandwidth, the overall performance is determined largely by the high-frequency components of the reference phase noise. In this region, the performance is independent of CR bandwidth, as can be seen by comparing the two panels in Figure \ref{fig:single_user_pll_bw}. 

    It is important to note that the reference noise at the the output sees a gain equal to the frequency multiplication ratio of the PLL. For RF operation ($<$6GHz) this ratio will typically be on the order of 10-100, while at mm-wave the multiplication ratio may exceed 1000. This means that even very high quality references may have very poor phase noise performance when referred to the PLL output frequency. For example, a 100 MHz reference with a -140 dBc/Hz noise floor multiplied to 60 GHz would have -85 dBc/Hz output referred reference noise from the PLL. In the simulated example, the receiver with this level of reference noise sees over 15 dB of improvement in SINR by moving to high bandwidth CR. This suggests that using high CR bandwidth is critical to achieve high-order modulation at mm-wave. This approach is taken in \cite{Okada:2013,Thomas:2015}.

%% file: sec-single_user_beam_scramble.tex
\section{Self Interference from Uncorrelated Phase Noise}
When analyzing a single-user array, it is important to consider the interaction of uncorrelated phase noise with the beamforming loops that are responsible for steering the beam. The user transmits signal $x$ through the wireless channel which is represented by $M\times1$ matrix $H$. Additionally, the receive signal is corrupted by AWGN source, $n$. The receiver uses $1\times M$ beamforming matrix $W$ to reconstruct the desired signal. $W$ is selected such that the resulting signal
\begin{equation}
\hat{x}=W(Hx+n)
\end{equation}
will reproduce $x$ with maximal signal-to-noise ratio (SNR). In the case of a single-user array, the optimal $W$ is simply
\begin{equation}
W = \hat{H}^{H}
\end{equation}
where $\hat{H}$ is the channel estimate and $^{H}$ denotes conjugate transpose.

The phase noise at each element can in general be expressed as a component $\phi_c$ which is correlated across all elements and a component $\phi_i$ which is uncorrelated and depends on element index $i$. Including phase noise, the beamformed signal will be
\begin{equation}
\hat{x} = W \mbox{\textbf{diag}} \{e^{j\phi_c+j\phi_i}\}(Hx+n)
\end{equation}
Here $\mbox{\textbf{diag}}\{\textbf{a}\}$ forms a diagonal matrix from vector $\textbf{a}$. Simplifying:
\begin{equation}
\label{eq:su-result}
\begin{split}
\hat{x} = e^{j\phi_c}W \mbox{\textbf{diag}} & \{e^{j\phi_i}\}Hx + Wn \\
= e^{j\phi_c} (\sum_{i=0}^{M-1}&{e^{j\phi_i}}) x + Wn
\end{split}
\end{equation}
since the statistics of $n$ are unchanged by the phase noise. It is instructive to express this as
\begin{equation}
\hat{x}[t]= e^{j\phi_c} \textbf{E}[e^{j\phi_i}]g[t]e^{j\theta[t]} x[t] + Wn
\end{equation}
where $t$ is the time index, $\textbf{E}[]$ denotes expectation, and the sum in (\ref{eq:su-result}) is represented as the product of its mean and time-varying residual gain and phase $g[t]$ and $\theta[t]$.

    This expression shows that correlated phase noise creates a time-varying rotation applied equally to all rows of $H$. Since this does not impact the relative weighting of different elements in the receiver, it is identical to phase noise in a single-element receiver. In contrast, the uncorrelated phase noise causes separate time-varying rotations to each row in H. This induces a time-varying discrepancy between the beamformer and the channel, which appears as gain variation in the receiver. This manifests as three main effects. First, there is a static gain loss represented by $\textbf{E}[e^{j\phi_i}]$. This arises from the signal filtering caused by partially incoherent addition across the array. Since this value is static, it can be estimated and compensated for by a slow automatic gain control (AGC) loop. Additionally, single-user beamforming is very robust to phase errors in the coefficients \cite{Bakr:2010}. Second, there are residual time-varying magnitude variations $g[t]$ which cannot be tracked and consequently appear as a gain error in the constellation plot\footnote{These gain variations could be tracked by a \textit{fast} AGC loop. However as discussed below this is only applicable to a single-user scenario.}. Finally, there is a residual phase noise $\theta[t]$ arising from the phase of the uncorrelated sum. The total phase noise at the output of the beamformer is $\phi_c+\theta[t]$; this phase noise can be filtered by the CR loop exactly as in the single-element case. As a result, following the CR loop the leftover noise consists of high-frequency phase noise and the gain self-interference. While the residual high-frequency phase noise is also present in a single-element receiver, the presence of gain self-interference is unique to array based systems and arises specifically from uncorrelated phase noise.

\begin{figure}[!t]
\centering
\includegraphics[width=3.4in]{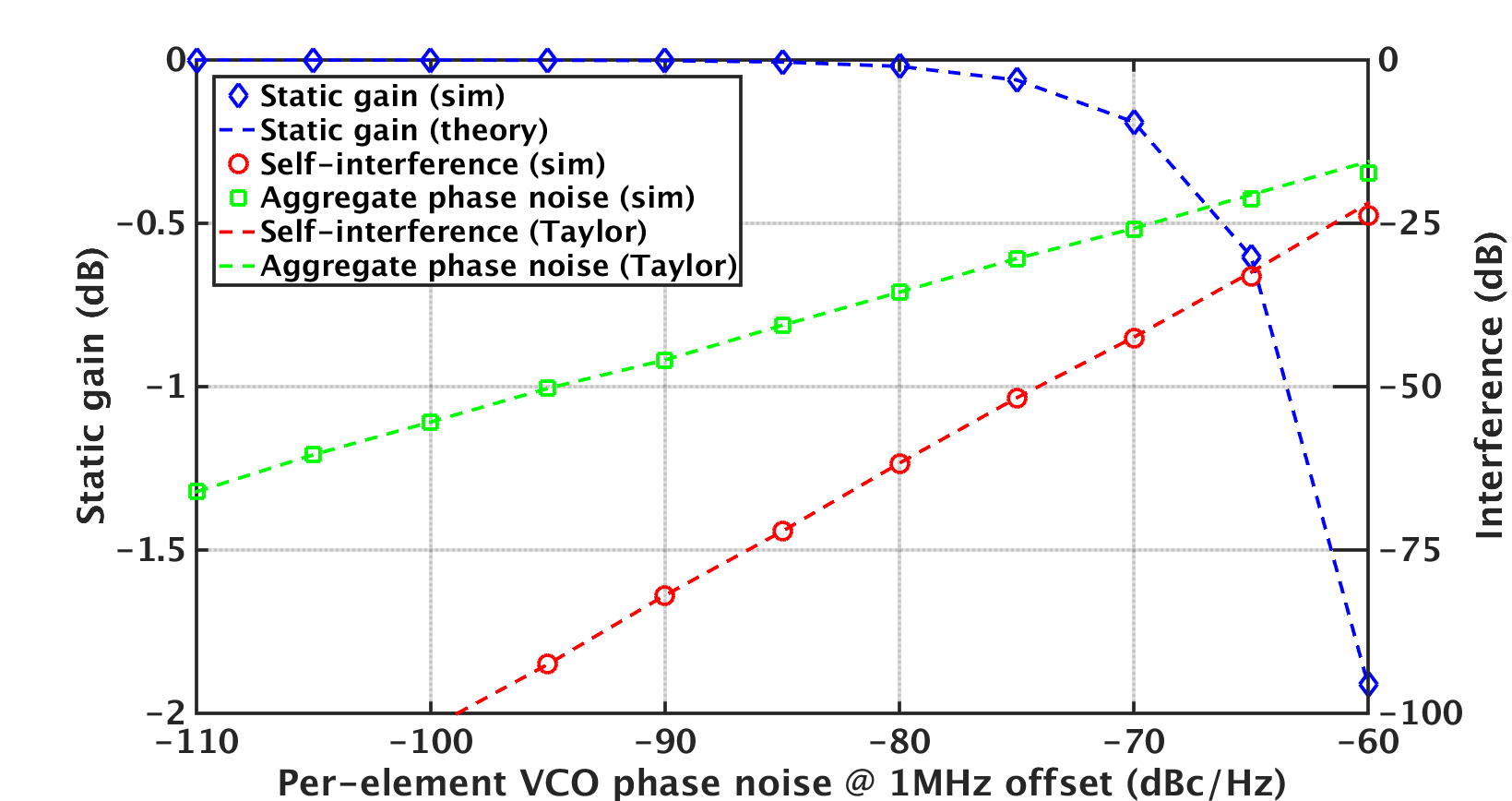}
\caption{Characterization of uncorrelated VCO phase noise for various VCO phase noise levels in a 16-element array. This figure shows static gain loss, self-interference, and net phase noise from Monte Carlo simulations as well as analytical model.}
\label{fig:ici_vs_pn}
\end{figure}
    
    The static gain error can be obtained using the characteristic function of the phase noise random process \cite{Petrovic:2007} and is given by
\begin{equation}
\textbf{E}[e^{j\phi_i}] = e^{-\sigma_{\phi}^{2}}
\end{equation}
where $\sigma_{\phi}^2$ is the variance of the phase noise. The gain and phase errors can be analyzed by considering the Taylor expansion
\begin{equation}
\sum_{i=0}^{M-1}{e^{j\phi_i}} \approx \sum_{i=0}^{M-1}{(1-\frac{1}{2}\phi^{2}_i)} + j\sum_{i=0}^{M-1}{\phi_i}
\end{equation}
This approximation is valid for small levels of phase noise. In this regime it is clear that the imaginary part is much larger than the real part (due to the linear compared to quadratic dependence on the phase). Consequently for small levels of phase noise
\begin{equation}
\begin{split}
\theta[t] \approx &\sum_{i=0}^{M-1}{\phi_i} \\
g[t] \approx \sum_{i=0}^{M-1}&{(1-\frac{1}{2}\phi^{2}_i)}
\end{split}
\end{equation}
For large phase noise variance it is difficult to obtain analytical expressions. Instead, we conduct Monte Carlo simulations to empirically obtain the distribution of the gain error. Figure \ref{fig:ici_vs_pn} shows the static gain, gain variations, and residual phase noise as a function of the mm-wave VCO's phase noise level, for $M=16$ and 5MHz PLL bandwidth. The theoretical analysis for the static gain error and the Taylor approximations for gain and phase error match the simulations very well. The SINR ceiling for this scenario is set by the difference between the static gain error and the gain self-interference. For high-order modulation schemes such as 64-QAM, where in excess of 25dB aggregate SINR is required, this self-interference contribution may pose a limiting factor on the achievable spectral efficiency.

\subsection{Impact of LO Architecture}

	The relative power of correlated and uncorrelated phase noise components, as well as their PSDs, is determined by the LO generation architecture of the array. For the CCG scheme, all the reference and VCO phase noise is correlated at every element. For the LCG scheme, the reference phase noise is correlated at every element while the VCO phase noise is fully uncorrelated. Furthermore, due to the phase noise filtering property of PLLs, in the LCG scheme the correlated phase noise will appear below the PLL bandwidth while the uncorrelated phase noise lies above it.

\begin{figure}[!t]
\centering
\includegraphics[width=3.5in]{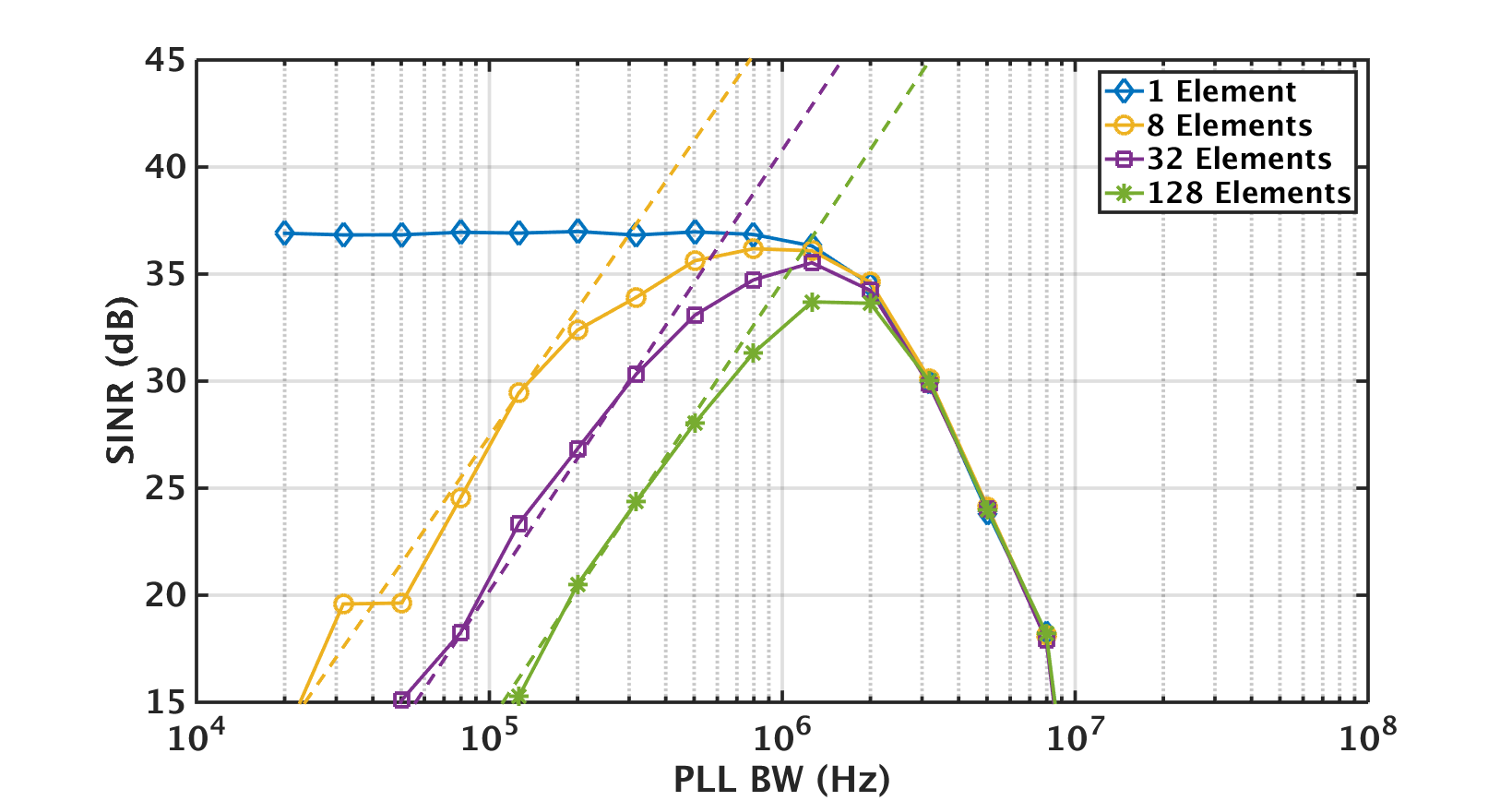}
\caption{SINR vs PLL bandwidth for a single-user array with local PLL scheme. Carrier recovery bandwidth is 10MHz, reference noise is -140dBc/Hz and VCO effective phase noise is -90dBc/Hz at 1MHz offset. Dashed lines indicate the predictions from the analytical model in the VCO-dominated regime.}
\label{fig:scramble_v_pllbw}
\end{figure}
    
    To accurately compare these two schemes, we consider a total VCO power budget which is identical for both scenarios. Consequently, the central VCO achieves a certain phase noise target $PN_0$ while each local VCO has $M$ times worse phase noise $PN_0 + 10\log_{10}M$. These scenarios are compared in Figure \ref{fig:scramble_v_pllbw}. These simulations all use the same 10 MHz CR bandwidth, -140 dBc/Hz noise floor reference, and 75 GHz local VCOs with ($-90 + 10 \log_{10}M$) dBc/Hz noise at 1 MHz offset. For high PLL bandwidths the phase noise is dominated by the reference and all systems have comparable performance. In contrast, for low PLL bandwidths the performance is limited by the VCO and specifically the self-interference generated by the uncorrelated phase noise. Because the VCO performance is deliberately degraded with the number of elements, the level of self-interference is increased for large array sizes. This manifests as degraded array-level SINR for low PLL bandwidths. These results in the VCO-dominated regime match very well with the prediction obtained from the simple analytical model. 
    
    A key observation is that for a system with an LCG scheme, the self-interference creates a lower bound of the optimal PLL bandwidth while the high reference noise sets an upper bound. This lower bound increases continually as the number of elements grows and the individual VCO performance is relaxed proportionally. Consequently, there exists an array size where these two bounds meet and the performance of the LCG scheme can no longer match that of the CCG scheme. In Figure \ref{fig:scramble_v_pllbw} this is observed for 128-element array, showing that the LCG scheme suffers a fundamental penalty for this design point.

\begin{figure}[!t]
\centering
\subfloat[Block diagram of array with IF-PLL based reference distribution.]{\includegraphics[width=2in]{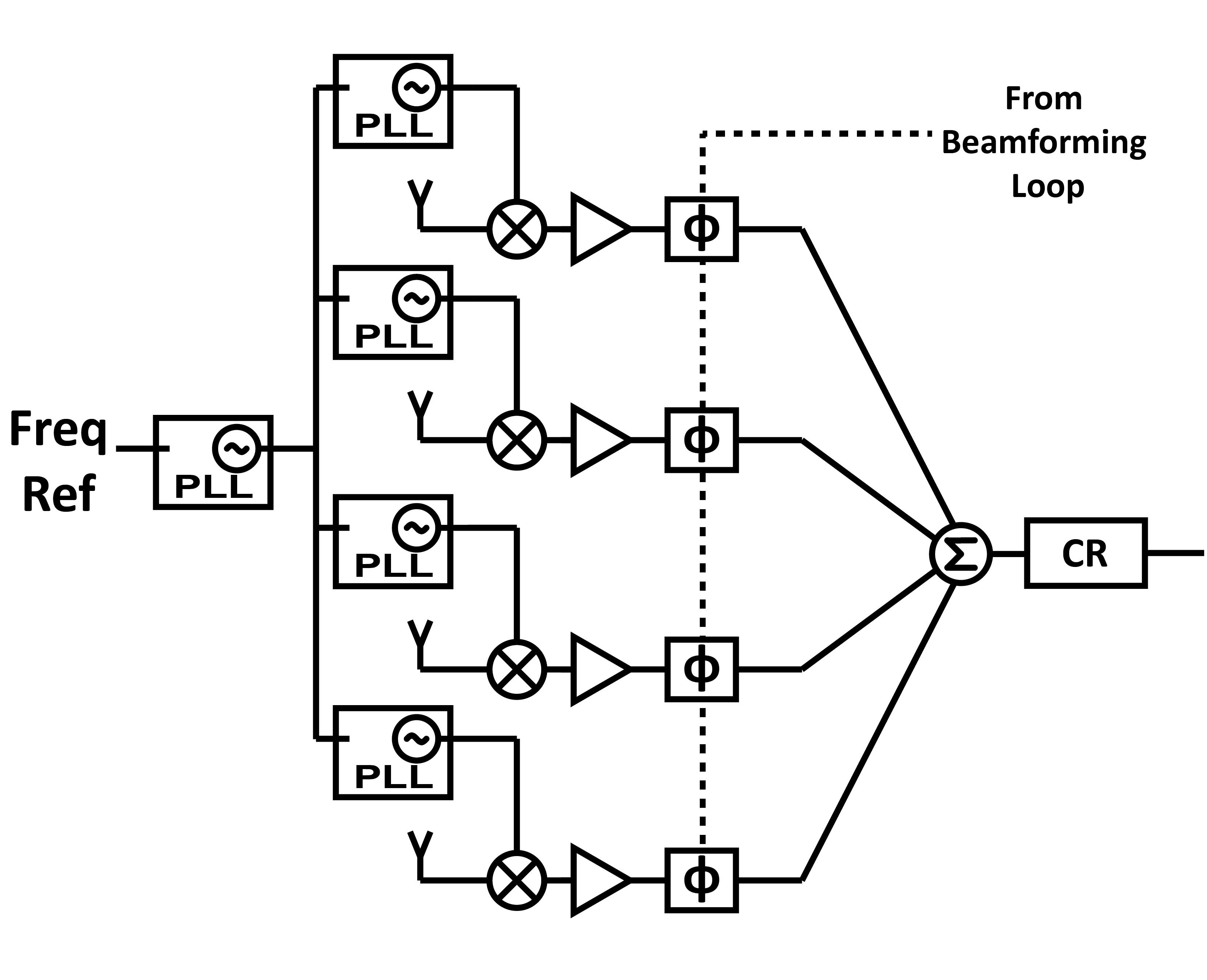}
\label{fig:ifpll_block}}

\subfloat[Phase noise PSD components at the output of the IF-PLL, referred to 75 GHz. The IF-PLL design is similar to \cite{Gao:2015}]{\includegraphics[width=3in]{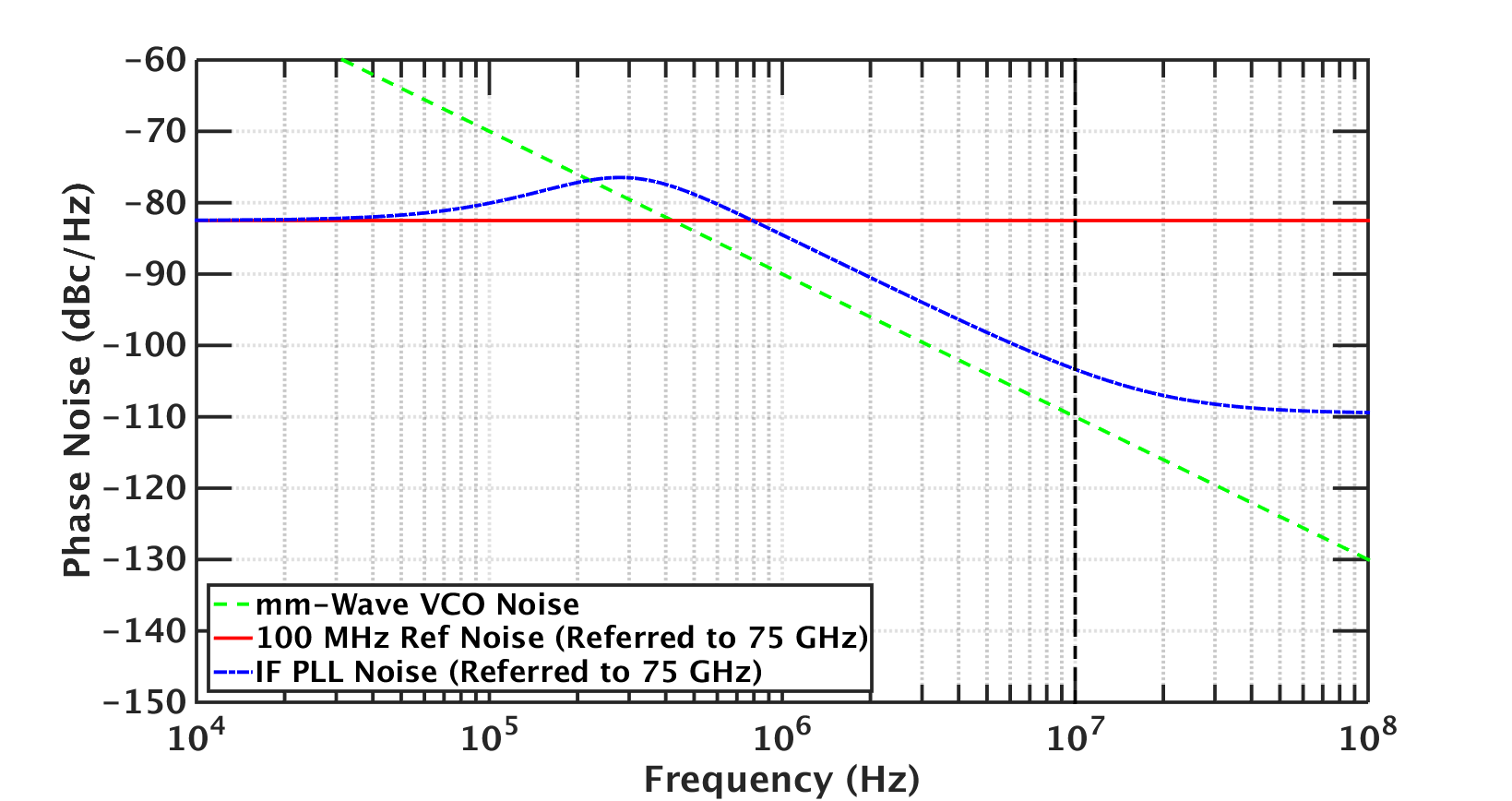}
\label{fig:ifpll_pn}}
\caption{IF-PLL LO generation and resulting phase noise filtering.}
\label{fig:IF_PLL_description}
\end{figure}

\subsection{Intermediate-frequency PLL}
	
    In practical deployments, the frequency of the reference is limited to be on the order of a few hundred MHz by the manufacturability of high resonance frequency crystals. It is the multiplication of these low frequency references to mm-wave (by a factor of over 100x) that sets the upper limit on PLL bandwidth and the potentially steep bandwidth optimum for large arrays using LCG. However, just because these low-frequency crystals are required to set the absolute frequency reference, they do not necessarily have to set the frequency distributed to the mm-wave PLLs. A PLL can multiply the low-frequency reference to an intermediate frequency (IF) on the order of low GHz which is distributed across the array (Figure \ref{fig:ifpll_block}). Because this frequency is still low relative to the carrier, IF distribution also consumes little power relative to the VCOs. 	
    
    This IF PLL acts as a "jitter cleaner" PLL. As shown in Figure \ref{fig:ifpll_pn}, the IF PLL filters higher-frequency components of the reference noise. As a consequence, the amount of reference noise propagated to the carrier is significantly reduced, allowing the bandwidth of the mm-wave PLLs to increase with little performance loss. 
    
\begin{figure}[!t]
\centering
\includegraphics[width=3.5in]{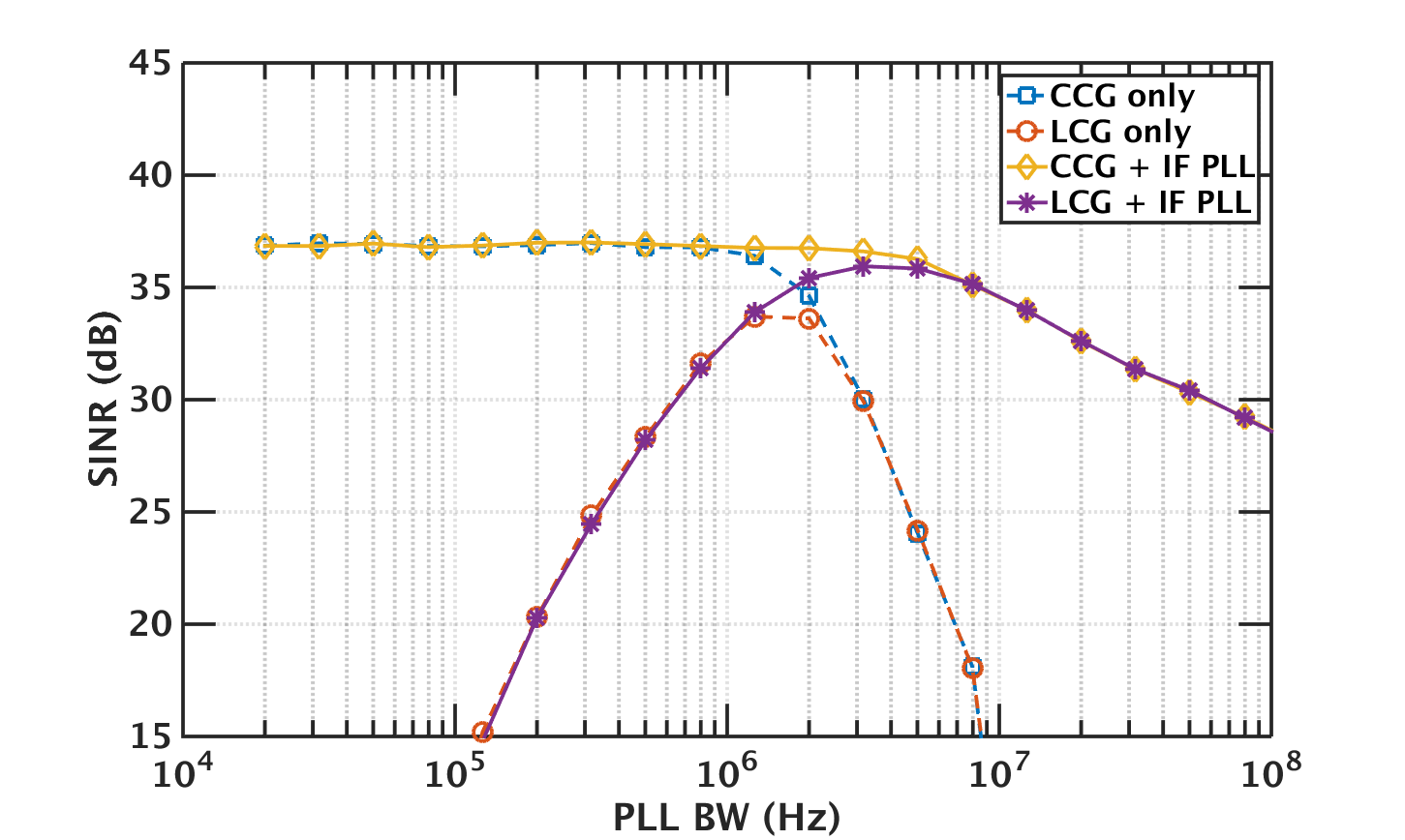}
\caption{Comparison of CCG and LCG schemes, both with and without IF-PLL reference distribution. Carrier recovery bandwidth is 10 MHz, reference noise is -140 dBc/Hz and VCO effective phase noise is -90dBc/Hz at 1 MHz offset, while a 5 GHz distribution is used for the IF-PLL case.}
\label{fig:if_pll_effect}
\end{figure}

    Figure \ref{fig:if_pll_effect} shows the performance of a 128-element array with several different distribution schemes. The same settings of 10 MHz CR bandwidth, -140 dBc/Hz noise floor reference, and 75 GHz VCOs with ($-90 + 10 \log_{10}M$) dBc/Hz noise at 1 MHz offset are used. The IF PLL uses a 5 GHz VCO with -110 dBc/Hz noise at 1 MHz offset and a 300 KHz PLL bandwidth. The 5 GHz distribution buffers have a -135 dBc/ Hz noise floor. This IF PLL is comparable to what is used in 802.11n transceivers \cite{Gao:2015}. When the 100 MHz reference is directly converted to mm-wave, the LCG scheme has a very steep optimum in PLL bandwidth and there is a fundamental performance loss compared to the CCG scheme. In contrast, the addition of an IF PLL is able to significantly broaden this optimum by reducing the amount of reference noise above the CR bandwidth when high PLL bandwidths are used. By reducing the penalty for using high bandwidth mm-wave PLLs, the performance gap between the LCG and CCG schemes is reduced.
    
%% file: sec-multi_user.tex
\section{Multi-User Interference from Uncorrelated Phase Noise}
The primary motivation for utilizing arrays at mm-wave is to compensate for the demanding link budget. However, the inherent directivity of the array also lends itself to spatial multiplexing which enables a significant increase in spectral efficiency. This high level of spatial multiplexing aligns with the vision of a massive MIMO system where a large number of antennas are deployed with closed-loop beamforming to serve many users with minimal inter-user interference. 

This multi-user operation can be considered as an extension of the previously outlined beamforming system where a system serving $K$ users now has a $M\times K$ channel matrix $H$ and the receiver implements a $K\times M$ beamforming matrix $W$. Many options for the selection of $W$ exist with varying trade-offs, but we will focus on a zero-forcing (ZF) beamformer, which suppresses inter-user interference by implementing
\begin{equation}
W = (\hat{H}^{H}\hat{H})^{-1}\hat{H}^{H}
\end{equation}
It is useful to consider this as a conjugate beamformer, $\hat{H}^H$, followed by a zero-forcing matrix $(\hat{H}^{H}\hat{H})^{-1}$. Thus the effective $K \times K$ channel before ZF is given by $A=\hat{H}^{H}H$.

With a ZF receiver, the received signal in the presence of correlated and uncorrelated phase noise is
\begin{equation}
\begin{split}
\label{eq:mu-pn}
\hat{x} = W \mbox{\textbf{diag}}\{&e^{j\phi_c+j\phi_i}\}(Hx+n) \\
= e^{j\phi_c}(\hat{H}^{H}\hat{H})^{-1}&\hat{H}^{H} \mbox{\textbf{diag}}\{e^{j\phi_i}\}Hx + Wn
\end{split}
\end{equation}
The effective channel in the presence of phase noise is
\begin{equation}
\begin{split}
A = \hat{H}^{H} \mbox{\textbf{diag}}& \{e^{j\phi_i}\}H \\
A_{ii} = &\sum_{i=0}^{M-1}{e^{\phi_i}} \\
A_{lk} = \sum_{i=0}^{M-1}&{H_{il}H_{ik}^{*}e^{\phi_i}}
\end{split}
\end{equation}
The diagonal elements $A_{ii}$ are identical to the single user case; as such, the same self-interference effects are present. In addition, the multi-user interaction must be considered. As can be seen in the off-diagonal terms, the uncorrelated phase noise causes a time-varying channel drift. Because this phase noise-corrupted channel is mismatched to the static ZF beamforming matrix, the ZF is unable to fully cancel inter-user interference, leading to a residual phase noise-induced zero-forcing error\footnotemark. 

\footnotetext{While the self-interference could be tracked by a fast AGC loop, tracking the inter-user interference would require a fast beamforming loop that tracks the effective channel $A$ and adapts the ZF matrix accordingly. This is computationally quite intensive.}
    
\begin{figure}[!t]
\centering
\includegraphics[width=3.5in]{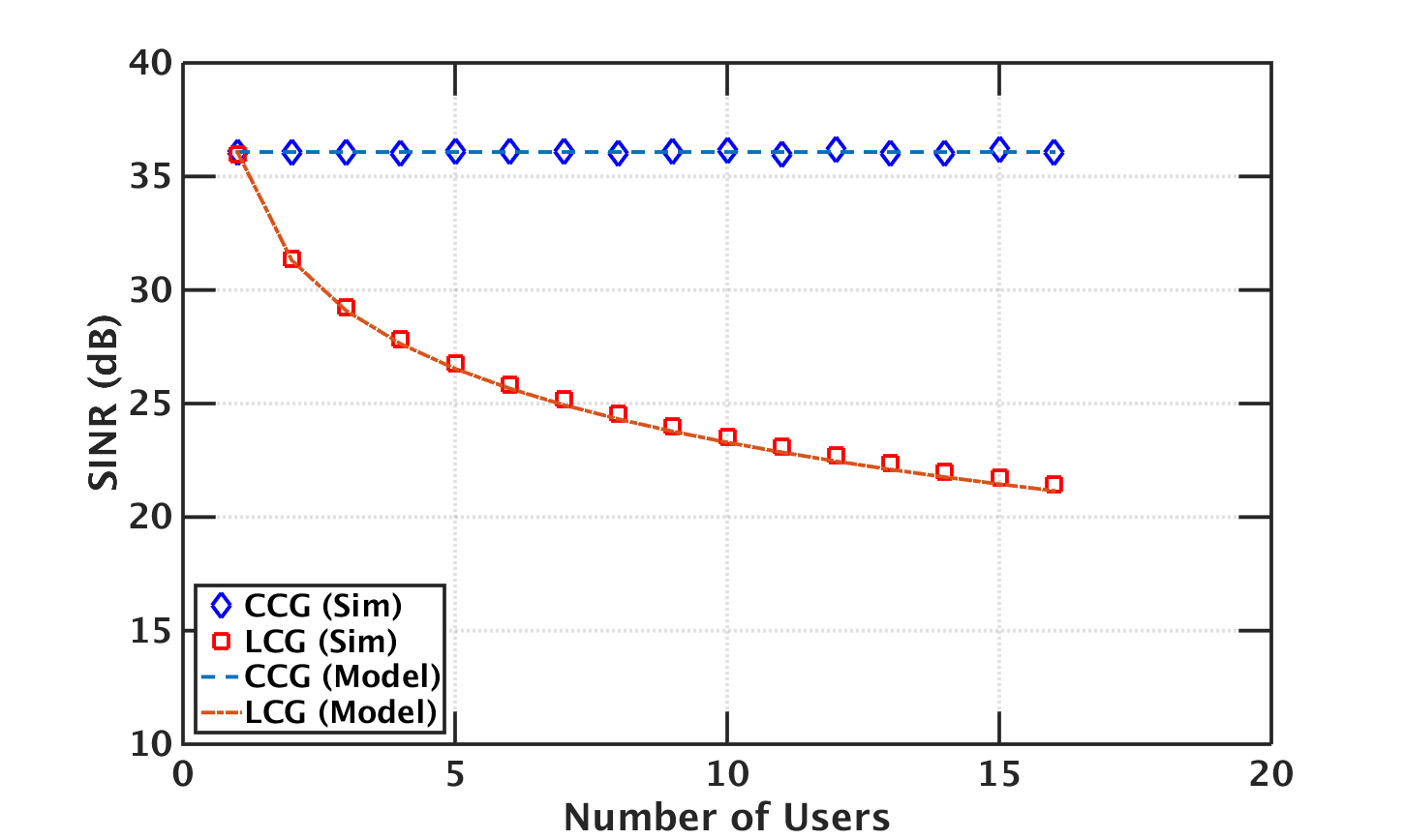}
\caption{Average SINR vs number of users for 128-element array with CCG or LCG scheme. Carrier recovery bandwidth is 10 MHz and PLL bandwidth is 5 MHz.}
\label{fig:SINR_v_users}
\end{figure}
    
    The multi-user SINR depends on the channel structure, particularly the underlying correlation of the different users' channels. We capture these effects by modeling the SINR as:
\begin{equation}
SINR = \frac{S_u}{N_t + S_u N_p + \alpha \gamma \sum_{j=1}{K-1}{S_j N_p}}
\end{equation}
where $S_u$ is the signal power, $N_t$ is the thermal noise power, $N_p$ is the phase noise power, and $S_j$ the power of the $j$'th user. If uplink power control is applied this becomes simply 
\begin{equation}
\label{eq:mu-sinr-model}
SINR = \frac{S_u}{N_t + S_u N_p + \alpha \gamma (K-1)S_u N_p}
\end{equation}
The model parameters $\alpha$ and $\gamma$ capture important effects. The parameter $\alpha$ describes the ratio of self- to inter-user interference power and depends on several factors including phase noise levels, the correlation between user channels, and the distributions of self-interference and zero-forcing errors. $\gamma$ is a purely architecture-dependent parameter which describes how the level of uncorrelated VCO phase noise depends on the array architecture. For LCG schemes, $\gamma=1$ since all VCO phase noise is uncorrelated. For CCG schemes, $\gamma=0$ because all VCO phase noise is correlated. For GCG, $\gamma$ assumes an intermediate value that is a function of $N$. 

	Figure \ref{fig:SINR_v_users} compares the performance of CCG and LCG using IF-PLL distribution. The mm-wave PLL bandwidth is 5 MHz in both scenarios. In this example the users are placed in line-of-sight channels with 10 degrees of angular separation between them. As expected, the CCG schemes performance is independent of the number of users, while the LCG performance degrades as the number of users is increased. These results match exactly with the model (\ref{eq:mu-sinr-model}), using $\alpha = 2$ and $\gamma = 1$. 

\begin{figure}[!t]
\centering
\includegraphics[width=3.5in]{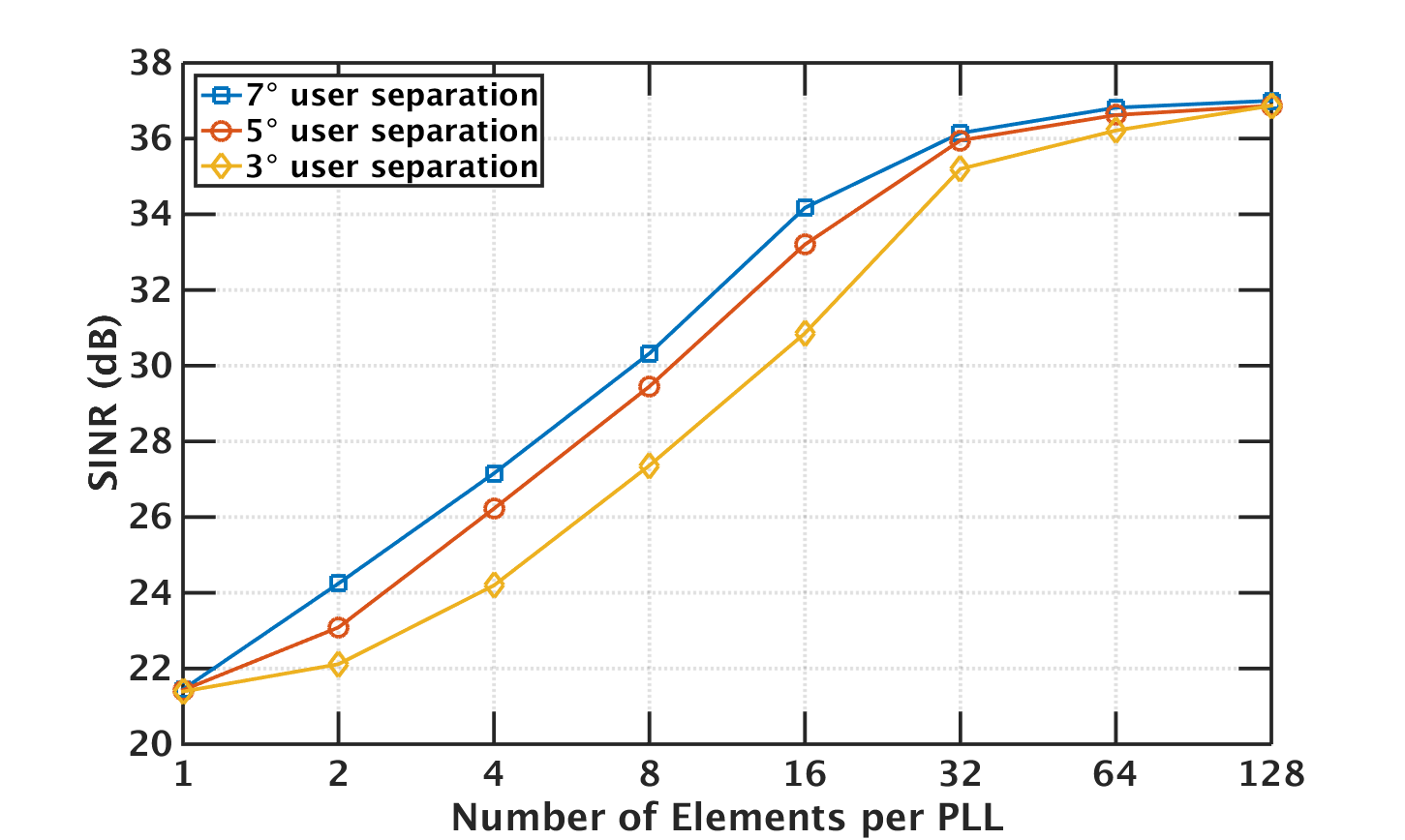}
\caption{Multi-user average SINR versus the number of elements per PLL in the GCG scheme. The array has 128 elements and there are 16 users.}
\label{fig:SINR_v_subarray}
\end{figure}

\subsection{Sharing PLL for Several Elements}

	As stated earlier, a CCG architecture is very unappealing for large scale arrays due to difficulty in distributing a mm-wave LO over such a large area. However, the CCG scheme fundamentally outperforms an LCG scheme in multi-user operation. In order to compromise between these two objectives, it is desired to keep some amount of correlation in the VCO noise in order to suppress inter-user interference. This can be achieved by using a GCG scheme. In the GCG scheme $N$ presents a design parameter that is available to influence $\gamma$ and therefore tune the dependence of the SINR on the number of users. 

	The set of elements that share a common PLL can be thought of as a subarray. Each subarray forms a collection of elements with fully correlated phase noise, consequently achieving some level of spatial filtering without any performance degradation from phase noise. When the signals from all subarrays are combined, the uncorrelated phase noise between subarrays tends to generate inter-user interference. However, as a result of the spatial filtering applied by the subarray, this inter-user interference is reduced and does not necessarily lead to large performance penalties. Figure \ref{fig:SINR_v_subarray} shows the receiver SINR when serving 16 users as a function of $N$. The dependence of the SINR on the angular separation of users illustrates the impact of the parameter $\alpha$, which is intended to capture channel dependence. This channel dependence arises due to increasing levels of demand on zero-forcing accuracy for channels with greater correlation. 
    
    The results indicate that as few as 32 elements per PLL can be used with little performance impact relative to a CCG scheme, making the LO distribution hardware more feasible. As $N$ is decreased beyond 32 the performance of the receiver steadily decreases.

 \subsection{Expected Throughput}
\begin{figure}[!t]
\centering
\includegraphics[width=3.5in]{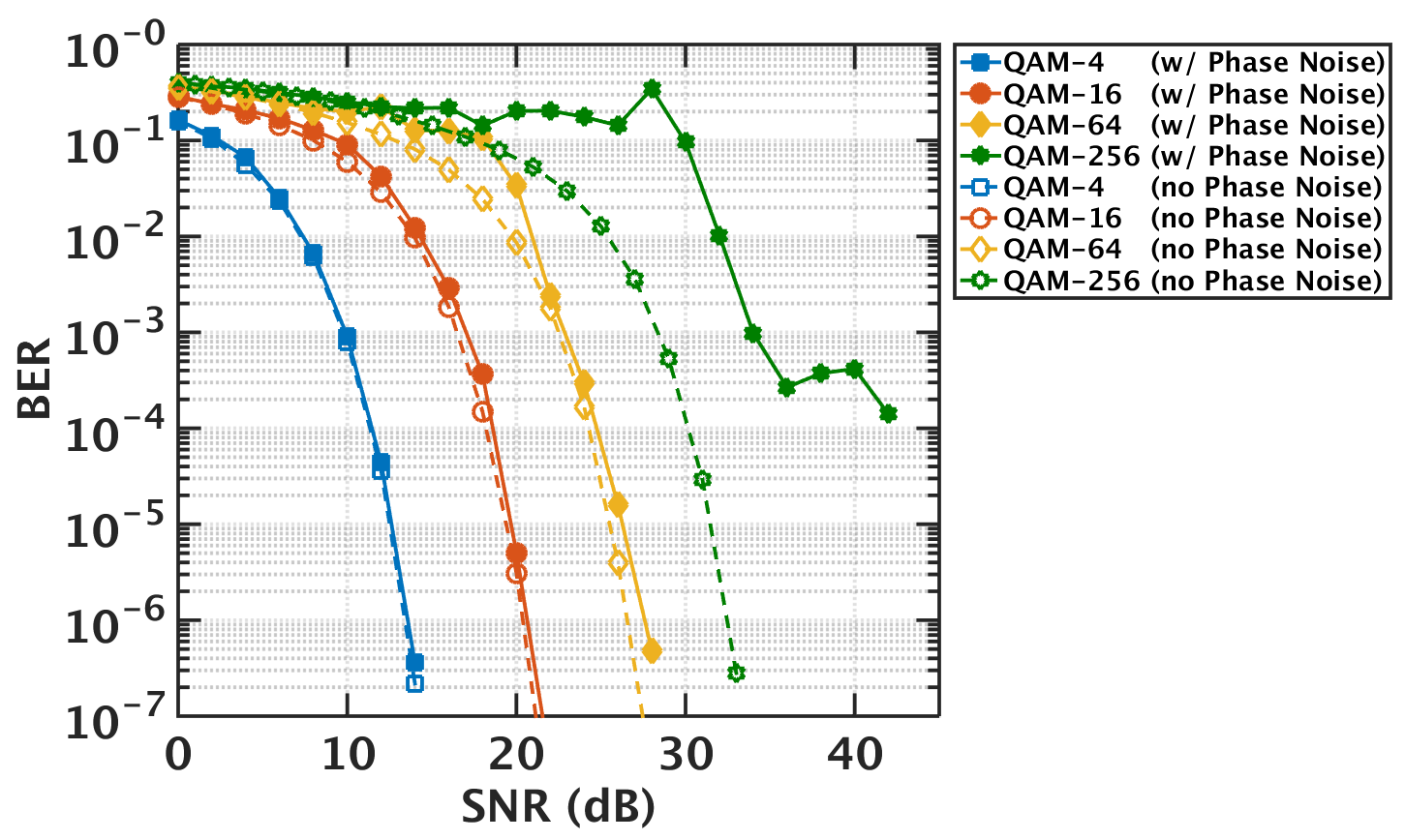}
\caption{Sum BER for 16 users versus thermal SNR with and without phase noise, for various constellation orders. The carrier recovery bandwidth is optimized for each thermal SNR level. The phase noise limited SINR is 36dB.}
\label{fig:BER_curves}
\end{figure}

	The previous sections have ignored all additional impairments to the receiver and only investigated the impact of phase noise. While it is difficult to include all of the expected non-idealities of a real transceiver chain, for completeness we have simulated the sum bit error rate (BER) of 16 users in the presence of AWGN from the channel for our proposed base station implementation from the previous section. The 4 VCOs used, each with -84 dBc/Hz of phase noise at 1 MHz offset, should be near-optimal in terms of energy efficiency and are in line with recently reported performance \cite{Sadhu:2015}. The BERs are shown in Figure \ref{fig:BER_curves} for different constellation orders as a function of the thermal SNR.
    
    By incorporating the proposed system architecture and design optimizations, the LO chain achieves excellent performance for QAM constellations of 4, 16, and 64. The phase noise-limited SINR is 36 dB, which is also sufficient to reach a BER floor of $5\times10^{-4}$ using 256-QAM. Since the decision-directed PLL can suffer from large bursts of errors when the BER is too large, the CR bandwidth is reduced at lower SNRs when the strength of the AWGN masks the impact of the phase noise. In a practical scenario this could easily be implemented by measuring the received signal strength of the channel estimation pilots, which typically use robust BPSK modulation. This effect accounts for the performance deviation observed at high BERs in the 64- and 256-QAM cases; in this regime other carrier recovery algorithms (e.g. a squaring loop) are more suitable than a decision-directed PLL. 

%% file: sec-conclusion.tex
\section{Conclusion}

In this paper, we have analyzed the system and circuit design trade-offs in the LO generation and distribution for mm-wave arrays. We have shown that the LO generation architecture strongly influences the power consumption and implementation complexity as well as the achievable performance. To properly understand the impact of the LO architecture, it is important to carefully model the interactions between multiple control loops and the effects of uncorrelated and correlated noise sources. 

We demonstrate the impact of these architecture choices via simulations of a 75 GHz system utilizing single carrier modulation and 2 GHz of channel bandwidth. While the optimum number of elements per PLL and optimum loop bandwidths may change for system realizations with different carrier frequency and channel bandwidth specifications, the overall trends and guidelines remain applicable. Additionally, many of the mechanisms that lead to our guidelines, such as self interference from gain errors and inter-user interference from zero-forcing error, which arise from uncorrelated phase noise, also remain relevant in OFDM systems. OFDM may add additional complexities due to inter-carrier interactions, which will require careful selection of sub-carrier bandwidth.

	The design guidelines obtained from our analysis are:
\begin{itemize}
\item{High carrier recovery bandwidths should be utilized to compensate for the poor quality of crystal references when multiplied up to mm-wave frequencies.}
\item{In single-user arrays, uncorrelated phase noise generates self-interference via gain errors in addition to phase noise. The self-interference can be managed by optimizing the PLL bandwidth to trade off VCO and reference noise.}
\item{In multi-user arrays, uncorrelated phase noise generates zero-forcing error due to the time-varying phase noise channel. This zero-forcing error can only be managed by applying some spatial filtering prior to introduction of uncorrelated phase noise. This requires sharing a single PLL between multiple array elements.}
\item{To balance the LO chain routing power, element synchronization requirements, and the self- and inter-user interference, the optimal LO chain architecture consists of a central IF-PLL along with subarray-based multiplication up to the mm-wave carrier frequency.}
\end{itemize}

Our system simulations indicate that, with the design insights identified here, we are able to achieve 16-way spatial multiplexing and support SINRs sufficient for 64-QAM modulation over 2 GHz of bandwidth. The results also suggest that 256-QAM modulation can be achieved with BERs below  $5\times10^{-4}$, although to achieve 256-QAM at mm-wave with spatial multiplexing, the impact of other transceiver non-idealities such as PA non-linearity, IQ imbalance, thermal/quantization noise, and inter-symbol interference must also be reduced beyond the current state of the art. The use of arrays may also offer improvement to these impairments by carefully co-designing the system architecture, signal processing algorithms, and circuits. 
 

%% file: sec-appendix.tex
\appendix[LO Distribution Power Model]
	This appendix derives a power model accounting for distribution and PLL power, parametrized by the number of radios per PLL, $N$. 
    
	The distribution power can be split into a portion which comes from routing loss and a portion which comes from loss in power splitters. The routing loss corresponds to all the routing distance which is downstream of the PLL. For a rectangular H-tree this can be expressed as
\begin{equation}
L_{route} = \frac{1}{2}\sum_{s=0}^{\log_2N-1}{2^{s-\log_2M}}(D_X+D_Y)L_{mm}
\label{eq:routing-loss}
\end{equation}
where $D_X$ and $D_Y$ are the x- and y-dimensions of the array and $L_{mm}$ is the loss per millimeter. It can be seen that the routing loss is insensitive to $N$ for small $N$, but grows rapidly when $N$ approaches $M$; this is intuitive because the greatest length of routing occurs at the top of the tree. Similarly, the loss in an $P$-way splitter can be expressed as
\begin{equation}
L_{sep} = L_{split}\log_P N 
\label{eq:splitter-loss}
\end{equation}
where $L_{split}$ is the loss of a single $P$-way splitter. 

	The output power of a central VCO can be computed based on the array phase noise spec $\mathcal{L}(f_{\Delta})$, the VCO FoM, and the VCO's efficiency $\eta_{osc}$ (ratio of output power to DC power) from (\ref{eq:vco-fom}). The output power of a distributed VCO is then divided by $N$. The distribution network for this VCO must overcome the aggregate routing loss in order to deliver the target amount of power to the load. Expressed in dBm, the distribution power is therefore
\begin{equation}
\begin{split}
P_{distr} = P&_{VCO} -10\log_{10}N+ L_{sep} + L_{route} \\
&- 10\log_{10}(\eta_{osc} - \eta_{driver})
\label{eq:distribution-power}
\end{split}
\end{equation}
where $\eta_{driver}$ is the efficiency of the driver. Note that if all distribution losses were zero, this quantity would be independent of $N$ and would reflect only the ideal propagation of the VCO power through the lossless dividing network. Therefore, any dependence of $P_{distr}$ on $N$ reflects power consumption specifically arising from architecture-dependent routing loss. Finally, the total PLL overhead is simply $N$ times the overhead of a single PLL \cite{Marcu:2011}.

	In the power comparison shown in Figure \ref{fig:routing-loss}, the following parameter values are used, representing commercially available components or recent reports in the literature. $L_{mm} = 0.2dB/mm$, $L_{split} = 1.5$ for $P = 4$ \cite{splitter-datasheet}, and $M=128$ at 75GHz, giving an array of $64mm \times 32mm$. Both the oscillator and amplifier efficiency are assumed to be 20\% \cite{Chao:2016}, the phase noise target is -90 dBc/Hz at 1MHz offset with a VCO FoM of 180dBc/Hz \cite{Sadhu:2015}, and the PLL overhead is 2mW \cite{Huang:2015}. 

%% file: sec-Acknow.tex
% use section* for acknowledgment
\section*{Acknowledgment}

Special thanks to Chris Hull from Intel Corporation, and the BWRC students, staff, and sponsors.

%% file: sec-bib.tex
%\begin{thebibliography}{1}

%\bibitem{IEEEhowto:kopka}
%H.~Kopka and P.~W. Daly, \emph{A Guide to \LaTeX}, 3rd~ed.\hskip 1em plus
%  0.5em minus 0.4em\relax Harlow, England: Addison-Wesley,…
%\end{thebibliography}

\bibliographystyle{IEEEtran}
% argument is your BibTeX string definitions and bibliography database(s)
\bibliography{hydra_lo_paper}